\documentclass[%
 aip,
 jmp,%
 amsmath,amssymb,
 reprint,%
 superscriptaddress,
 longbibliography
]{revtex4-1}

\usepackage{bm}       
\usepackage{color,soul}  
\usepackage{dcolumn}  
\usepackage{graphicx} 
\usepackage{multirow}

\usepackage{titlesec}
\titlespacing*{\subsection}
{0pt}{2.5ex}{0.5ex}
\titlespacing*{\section}
{0pt}{3.5ex}{1.0ex}

\newcommand{\bra}[1]{\ensuremath{\langle \: #1 \: |}}
\newcommand{\ket}[1]{\ensuremath{| \: #1 \: \rangle}}
\newcommand{\w}{\ensuremath{\omega}}
\newcommand{\NMax}{\ensuremath{N_\text{max}}}
\newcolumntype{C}{>{$}c<{$}}

\begin{document}

\title{Vibrational Density Matrix Renormalization Group}

\author{Alberto Baiardi}
\affiliation{ 
Scuola Normale Superiore, Piazza dei Cavalieri 7, 56126 Pisa, Italy
}
\author{Christopher J. Stein}
\affiliation{ 
ETH Z\"urich, Laboratorium f\"ur Physikalische Chemie, Vladimir-Prelog-Weg 2, 8093 Z\"urich, Switzerland
}
\author{Vincenzo Barone}
\email[Corresponding author: ]{vincenzo.barone@sns.it}
\affiliation{ 
Scuola Normale Superiore, Piazza dei Cavalieri 7, 56126 Pisa, Italy
}
\author{Markus Reiher}
\email[Corresponding author: ]{markus.reiher@phys.chem.ethz.ch}
\affiliation{ 
ETH Z\"urich, Laboratorium f\"ur Physikalische Chemie, Vladimir-Prelog-Weg 2, 8093 Z\"urich, Switzerland
}

\date{\today}

\begin{abstract}
Variational approaches for the calculation of vibrational wave functions and energies are a natural route to obtain highly accurate results with controllable errors.
Here, we demonstrate how the density matrix renormalization group (DMRG) can be exploited to optimize vibrational wave functions (vDMRG) expressed as matrix product states.
We study the convergence of these calculations with respect to the size of the local basis of each mode, the number of renormalized block states, and the number of DMRG sweeps required.
We demonstrate the high accuracy achieved by vDMRG for small molecules that were intensively studied in the literature.
We then proceed to show that the complete fingerprint region of the sarcosyn-glycin dipeptide can be calculated with vDMRG.

\end{abstract}

\maketitle

\setlength{\parindent}{0cm}
\setlength{\parskip}{0.6em plus0.2em minus0.1em}

\section{Introduction}                                                                                      

Vibrational spectroscopy is a valuable tool for the characterization of molecular systems. 
Different techniques 
--- ranging from standard infrared absorption,\cite{IRBook} to more intricate spectroscopies such as vibrational circular dichroism,\cite{VCDBook} Raman,\cite{Nafie2014_RamanReview} and Raman Optical Activity\cite{Bour2014_ROAReview} ---
allow one to record vibrational spectra for a detailed characterization of chemical systems, that span a wide range
from molecules of astrochemical interest\cite{Barth2007_InfraredProtein,Herbst2009_ReviewAstro,Tielens2013_MolecularUniverse,Puzzarini2015_AccountAstro} to large biomolecules.\cite{Jilie2007_ReviewFTIR,Reiher2007_BioBook}
In order to decode the detailed information contained in an experimental spectrum of a complex system, simple selection rules based on semi-empirical Hamiltonians are insufficient, and \textit{ab initio} calculations are required.

Computational approaches for the calculation of vibrational properties of molecular systems can be assigned to two main classes, 
namely variational and perturbative approaches. 
In variational calculations, the vibrational energies and wave functions are obtained by diagonalization of a vibrational Hamiltonian in a given basis set such as the harmonic-oscillator eigenfunctions\cite{Whitehead1975_VCIFirst,Romanowski1985_VCIFormaldehyde,Carter1986_VariationalVibrations,Carbonniere2004_VCICoriolis,Crittenden2016_PyVCI} or eigenfunctions from a vibrational self-consistent field (VSCF) calculation.\cite{Carter1997_VCI,Gerber1999_VSCF,Gerber2002_CCVSCF,Hirata2007_CO2,Neff2009_LargeScaleVCI} 
The fraction of vibrational correlation energy that is missing in the VSCF mean-field approach can then be captured by vibrational configuration interaction\cite{chris07,scri08,Neff2009_LargeScaleVCI,stro11} or vibrational coupled cluster\cite{chris04,chris07} methods.
With variational approaches it is possible to obtain fully converged results for a given Hamiltonian by systematically increasing the basis set.
The Hamiltonian in turn depends on the electronic potential that is calculated with quantum-chemical \textit{ab initio}
electronic structure methods and remains the main source of error.

Unfortunately, the computational cost of these variational approaches grows exponentially with the size of the system and limits the range of application to rather small molecules (up to 20 atoms). 
Perturbative approaches on the other hand, such as the most commonly applied vibrational second-order perturbation theory (VPT2)\cite{Mills1961_VPT2First,gerb05,Barone2005_VPT2,Krasnoshchekov2014_VVPT2,Rosnik2014_VPT2K} and vibrational M{\o}ller-Plesset perturbation theory,\cite{Norris1996_VMP2First,Christiansen2003_VMP2,Changala2016_VMP2CurvCoord} are computationally more feasible and can be applied to systems with up to 100 atoms.
However, although reliable results can be obtained with VPT2 for semi-rigid systems, this approach certainly fails for molecules with shallow potential energy surfaces (PESs) and corresponding highly anharmonic, large-amplitude modes.

Several techniques have been developed and successfully applied to reduce the computational effort of variational approaches.
Basis pruning algorithms\cite{Dawes2005_BasisPruning,Avila2011_C2H4PrunedBasis}  were developed with the aim to include only a limited number of basis functions in the variational calculation. 
An alternative approach is based on so-called contracted basis techniques, in which basis functions are obtained by diagonalizing sub-blocks of the full Hamiltonian, involving only strongly interacting coordinates.\cite{Bramley1993_Contraction,Wang2002_ContractedBasis} 
The computational cost of variational calculations can also be reduced with local mode techniques\cite{Jacob2009_LocalModes,Jacob2009_PolypeptideLocalModes,Panek2014_LocalModes,Cheng2014_LocalModesVCI,klin15,pane16} instead of normal coordinates to reduce the number of off diagonal anharmonic couplings.
A further alternative, whose analog is widely applied in electronic structure theory, but hardly explored in vibrational calculations, is the parameterization of the vibrational wave function in tensor formats, such as canonical decomposition\cite{lecl16a} and matrix product states (MPS).
These wave function representations must then be optimized with efficient algorithms such as the density matrix renormalization group (DMRG)
algorithm\cite{whit92,whit93,Schollwoeck2005,lege08,chan08,chan09,mart10,mart11,chan11,Schollwoeck2011,kura14,wout14,yana15,szal15,knec16,chan16} for the optimization of MPS wave functions.
Here, we develop a DMRG optimization of MPS representations for vibrational wave functions and energies. 
We denote our approach as vDMRG. 
Recently, the eigenfunctions of a vibrational Hamiltonian were expressed in a tensor train format\cite{Rakhuba2016_TensorTrain}, with a discrete variable representation (DVR) basis set\cite{Colbert1992_DVR}. 
While the tensor train format is algebraically equivalent to the MPS format, the optimization protocol proposed in Ref.\ \onlinecite{Rakhuba2016_TensorTrain} is different from the two-site DMRG algorithm chosen for our vDMRG approach.
Moreover, our implementation of vDMRG expresses both the wave function and the Hamiltonian in tensor format, as an MPS and matrix product operator (MPO), respectively. 

Our paper is organized as follows. 
Section \ref{sec:theory} describes the underlying theory of vDMRG.
After this brief description of the computational details of the implementation, the application of vDMRG to several molecules of varying size is discussed. 
First, we demonstrate the reliability of vDMRG at the example of a triatomic molecule, {ClO$_2$}, for which fully converged variational energies can be 
easily calculated.
Then, two medium-sized molecules ({CH$_3$CN} and {C$_2$H$_4$}) are studied in detail. 
For {CH$_3$CN}, we chose a quartic PES from density functional theory calculations\cite{Carbonniere2005_CH3CN,lecl16a,Rakhuba2016_TensorTrain} and
for {C$_2$H$_4$} a sextic PES from accurate coupled cluster calculations.\cite{Delahaye2014_EthylenePES} 
Results are compared to experimental data.\cite{Georges1999_C2H4Exp}
Finally, the vibrational properties of the sarcosyn-glycin dipeptide ({SarGly$^+$}) are calculated to assess the reliability of vDMRG for large systems, for which standard variational calculations are generally unfeasible.

\section{vDMRG Theory}
\label{sec:theory}

The vibrational wave function $\ket{\Psi_k}$ of a molecule in the $k$-th vibrational state with $L$ degrees of freedom can be expressed by a 
full configuration interaction (FCI) expansion, 
\begin{equation}
  \ket{\Psi_k} = \sum_{\sigma_1,...,\sigma_L}  
                          {C}_{\sigma_1, \ldots , \sigma_\text{L}}^{(k)} \ket{\sigma_1,\ldots,\sigma_\text{L}},
  \label{eq:VCI_ansatz}
\end{equation} 
where the occupation number vectors are built from a basis of one-dimensional functions for each vibrational degree of freedom.
Whereas for an electronic wave function, due to Fermi-Dirac statistics, the occupation number of each orbital cannot exceed 1, 
it is unbounded for a mode in a bosonic vibrational wave function.
To limit the size of the basis set in Eq.~(\ref{eq:VCI_ansatz}), an upper bound ($N_\text{max}^\text{i}$) for the occupation number of each mode has to be defined.

For a total vibrational state $k$ expressed in terms of occupation number vectors $\ket{\sigma_1,...,\sigma_L}$, the MPS $N$-body wave function $\ket{\Psi_k}$ 
reads
\begin{equation}
  \ket{\Psi_k} = \sum_{\sigma_1,...,\sigma_L} \sum_{a_1,...,a_\text{L-1}}^m 
               {M}_{1,a_1}^{(k)\sigma_1} {M}_{a_1,a_2}^{(k)\sigma_2}...
               {M}_{a_\text{L-1},1}^{(k)\sigma_L}
               \ket{\sigma_1,...,\sigma_L}               
  \label{eq:DMRG_ansatz}
\end{equation}
expressed in a basis set of $L$ one-dimensional basis functions. 
The coefficients of the linear expansion in Eq.~(\ref{eq:DMRG_ansatz}) are decomposed as products of $L$ matrices 
$\bm{M}^{(k)\sigma_i}=\{{M}_{a_{i-1},a_i}^{(k)\sigma_i}\}$ with maximum dimension $m$ 
(with the exception of $\bm{M}^{(k)\sigma_1}$ and $\bm{M}^{(k)\sigma_L}$, which are row and column vectors, respectively, of that dimension). 
By restricting the maximum dimension of the individual matrices $\bm{M}^{(k)\sigma_i}$ to $m$, the DMRG algorithm achieves a reduction of the scaling from exponential to polynomial.
This is a significant computational advantage over the FCI expansion, where the exponential scaling\cite{aqui8} limits applications to small systems.

Hence, the CI coefficients are encoded in MPS form, 
\begin{equation}
  {C}_{\sigma_1, \ldots , \sigma_\text{L}}^{(k)} = \sum_{a_1, \ldots , a_{L-1}}^m 
  {M}_{1,a_1}^{(k) \sigma_1} {M}_{a_1,a_2}^{(k) \sigma_2} \dots
  {M}_{a_{L-1},1}^{(k) \sigma_L} \, ,
  \label{eq:DMRG_ansatz_vibrations}
\end{equation}
which is an approximation for finite choices of $m$, which we denote the 'number of renormalized block states'.
The product structure of Eq.~(\ref{eq:DMRG_ansatz_vibrations}) implies a one-dimensional ordering of the vibrational modes $i$ that are now associated with the matrices $\bm{M}^{(k)\sigma_i}$.
We refer to these modes as \textit{sites} and denote the sequence of these sites as a \textit{lattice} for consistency with the general DMRG nomenclature.

In vDMRG, the matrices $\bm{M}^{(k)\sigma_i}$ are calculated variationally by minimizing the expectation value of the vibrational Hamiltonian $\mathcal{H}_\text{vib}$ over the state $\ket{\Psi_k}$. In this work, 
an approximate form of the Watson Hamiltonian\cite{wats68}, in which only the second-order Coriolis terms are included,\cite{Carbonniere2004_VCICoriolis} and higher-order terms in the expansion of the inertia tensor are neglected,
  \begin{align}
\hspace*{-0.2cm}    \mathcal{H}_\text{vib} =
&   \frac{1}{2}  \sum_{i=1}^{L}    \omega_i \left( \hat{p}_i^2 + \hat{q}_i^2 \right) \nonumber\\
\hspace*{-0.2cm}            & + \frac{1}{6}  \sum_{ijk=1}^{L}  \boldsymbol{\Phi}_{ijk} \hat{q}_i \hat{q}_j \hat{q}_k +
\frac{1}{24} \sum_{ijkl=1}^{L} \boldsymbol{\Phi}_{ijkl} \hat{q}_i \hat{q}_j \hat{q}_k \hat{q}_l \nonumber\\
\hspace*{-0.2cm}            & +  \sum_{ijkl=1}^{L} \sum_{\tau}^{x,y,z} B^\tau
\zeta_{ij}^\tau \zeta_{kl}^\tau \sqrt{\frac{\w_j \w_l}{\w_i \w_k}} \hat{q}_i \hat{p}_j \hat{q}_k \hat{p}_l  ,
\label{eq:Watson_Hamiltonian}
\end{align}
is applied for the calculation of vibrational energies.
Here, $\hat{q}_i$ is the position operator associated with the $i$-th normal mode, and $\hat{p}_i$ is its conjugate momentum. 
In a fourth-order (quartic) Taylor expansion of the potential, $\omega_i$ are the harmonic frequencies and $\boldsymbol{\Phi}_{ijk}$ and $\boldsymbol{\Phi}_{ijkl}$ are the third- and fourth-order reduced force constants, respectively, which can be defined in terms of the third- and fourth-order partial derivatives, $k_{ijk}$ and $k_{ijkl}$, of the PES, 
\begin{equation}
  \boldsymbol{\Phi}_{ijk}  = \frac{k_{ijk}}{\sqrt{\w_i \w_j \w_k}} \quad \text{and}\quad
  \boldsymbol{\Phi}_{ijkl} = \frac{k_{ijkl}}{\sqrt{\w_i \w_j \w_k \w_l}}\, .
  \label{eq:def_constants}
\end{equation}
In Eq.~(\ref{eq:Watson_Hamiltonian}), $B^{\tau}$ are the rotational constants and $\zeta_{ij}^\tau$ the Coriolis coupling constants.
As the generalization of Eq.~(\ref{eq:Watson_Hamiltonian}) to support also higher-order terms is straightforward,
calculations with fifth- and sixth-order potentials are also presented in this work.

To exploit the DMRG formalism, a second-quantized Hamiltonian is required.
The second-quantized form of the vibrational Hamiltonian $\mathcal{H}_\text{vib}$ can be obtained by the following substitution:
\begin{align}
    \hat{p}_i &= \frac{1}{\sqrt{2}} \left( \hat{b}_i^+ - \hat{b}_i \right) \, , 
      \label{eq:SQOperators_Vibrations_1}
\end{align}    
\begin{align}    
    \hat{q}_i &= \frac{1}{\sqrt{2}} \left( \hat{b}_i^+ + \hat{b}_i \right) \, ,
  \label{eq:SQOperators_Vibrations_2}
\end{align}
where $\hat{b}_i$ and $\hat{b}_i^+$ are the bosonic annihilation and creation operators for the $i$-th vibrational mode, obeying the following rules:
\begin{align}
\hspace*{-0.2cm}
    \hat{b}_i^+ \ket{\sigma_1,...&,\sigma_i,...,\sigma_L} \nonumber\\
              &= \sqrt{\sigma_i+1} \ket{\sigma_1,...,\sigma_i+1,...,\sigma_L} ,\\
\hspace*{-0.2cm}
    \hat{b}_i   \ket{\sigma_1,...&,\sigma_i,...,\sigma_L} \nonumber\\
              &=
    \left\{\begin{array}{ll}
    \sqrt{\sigma_i}   \ket{\sigma_1,...,\sigma_i-1,...,\sigma_L} & \text{if} \, \sigma_i > 0\\
    0  & \text{if} \,  \sigma_i = 0
    \end{array}\right. .
  \label{eq:SQOperator_Vibrations_2}
\end{align}

The second-quantized form of the Watson Hamiltonian can be obtained by substitution of Eqs.~(\ref{eq:SQOperators_Vibrations_1}) and (\ref{eq:SQOperators_Vibrations_2}) in Eq.~(\ref{eq:Watson_Hamiltonian}) and reads:\cite{Hirata2014_SecondQuantization}. 
 \begin{align}
		& \mathcal{H}_\text{vib} = \sum_{i=1}^{L} \w_i \left( \hat{b}_i^+\hat{b}_i + \frac{1}{2} \right) \nonumber\\
		& + \frac{1}{12\sqrt{2}} \sum_{ijk=1}^{L} \boldsymbol{\Phi}_{ijk}
		( \hat{b}_i^+ + \hat{b}_i ) ( \hat{b}_j^+ + \hat{b}_j ) ( \hat{b}_k^+ + \hat{b}_k ) \nonumber\\
		& + \frac{1}{96} \sum_{ijkl=1}^{L} \boldsymbol{\Phi}_{ijkl}
		( \hat{b}_i^+ + \hat{b}_i ) ( \hat{b}_j^+ + \hat{b}_j ) ( \hat{b}_k^+ + \hat{b}_k )
		( \hat{b}_l^+ + \hat{b}_l ) \nonumber\\
		& + \frac{1}{4} \sum_{ijkl=1}^{L} \sum_{\tau}^{x,y,z} B^\tau
		\zeta_{ij}^\tau \zeta_{kl}^\tau \sqrt{\frac{\w_j \w_l}{\w_i \w_k}}
		( \hat{b}_i^+ + \hat{b}_i ) ( \hat{b}_j^+ - \hat{b}_j ) \nonumber\\
		& \hspace{4cm} \times ( \hat{b}_k^+ + \hat{b}_k ) ( \hat{b}_l^+ - \hat{b}_l )
  \label{eq:WatsonHamiltonian}
  \end{align}
Hence, the third-order potential term, for example, can be written as a sum of all possible products of bosonic
creation or annihilation operators localized on sites $i$, $j$, and $k$ (such as $\hat{b}_i \hat{b}_j^+ \hat{b}_k$). 
A similar result is obtained for the fourth-order potential term, but in this case 16 terms are present.
More attention must be paid to the evaluation of the sign associated to a Coriolis term due to the presence of the momentum operator. 
At variance with the electronic Hamiltonian, the second-quantized form of the vibrational Hamiltonian contains operator strings with different 
numbers of creation and annihilation operators and, as a consequence, a computational scheme such as DMRG that generates explicit matrix representations of these elementary operators, will face significant challenges that can only be well met by a strictly modular, general, and object-oriented implementation.

An operator $\hat{W}$ can be written as an MPO,
\begin{equation}
\hat{W} = \sum_{\bm{\sigma\sigma'}} \sum_{b_1,\dots,b_{L-1}} W_{1 b_1}^{\sigma_1,\sigma_1'} \dots W_{b_{l-1} b_l}^{\sigma_l,\sigma_l'} \dots W_{b_{L-1} 1}^{\sigma_L,\sigma_L'} \ket{\bm{\sigma}} \bra{\bm{\sigma'}}\, ,
\label{mpo_general}
\end{equation}
where $\ket{\bm{\sigma}}$ is a compact notation for $\ket{\sigma_1,\ldots,\sigma_\text{L}}$.
Compared to the MPS matrices in Eq.~(\ref{eq:DMRG_ansatz_vibrations}), the MPO matrices $\bm{W}^{\sigma_i,\sigma_i'}$ 
(the summation over the $b_i$ indices is equivalent to a matrix-matrix multiplication) have two superscripts that specify 
elements of the bra-ket vectors $\ket{\bm{\sigma}} \bra{\bm{\sigma'}}$.
While the dimension of the MPS matrices $\bm{M}^{(k)\sigma_i}$ is adapted to a maximum value during the DMRG optimization, 
the dimensions of the MPO matrices $\bm{W}^{\sigma_i,\sigma_i'}$, and therefore the $b_i$ indices, are fixed for each site by the 
particular form of the operator encoded in MPO format.

The MPO can now be constructed from matrix representations of the elementary operators $\hat{b}_i^+$ and $\hat{b}_i$.
To understand the formulation of the Watson Hamiltonian in Eq.~(\ref{eq:WatsonHamiltonian}) in MPO format, we write explicitly the construction of the first (harmonic) term for mode $i$ that we denote $\hat{W}_{\text{harm.}}^\text{Watson}(i)$:
\begin{equation}
\hat{W}_{\text{harm.}}^\text{Watson}(i) =  \omega_i \cdot \hat{I}_1 \otimes \dots \otimes \hat{b}_i^+ \cdot \hat{b}_i \otimes \dots \otimes \hat{I}_L \, ,
\end{equation}
with unit matrices $\hat{I}$ on all sites but $i$.
Therefore, the construction of the Hamiltonian scales with $L^4$ for quartic force fields 
(as in Eq.~(\ref{eq:WatsonHamiltonian})) and $L^6$ for sextic force fields in a naive MPO construction.

In the DMRG algorithm, the MPS optimization requires the repeated application of exact diagonalization and singular value decomposition.
This procedure is called a sweep algorithm because the MPS site matrices are subsequently optimized from the first to the $L$-th site 
(one sweep) before they are optimized in reverse order from the $L$-th to the first site (next sweep).
This is an optimization of reduced dimensional basis states contributing to the quantum state in a least-squares sense.
It is a peculiarity of the algorithm that parts of the total vibrational MPS wave function, entering the energy expectation value, 
are not known in the first sweep.
Therefore, an initial guess with random numbers as MPS entries or an MPS solution of an MPO with only harmonic contributions can be applied.

While a certain excitation rank has to be fixed in standard VCI calculations, this is not the case for vDMRG. vDMRG approximates the correct wave function in a least-squares sense and optimizes the coefficients of the configurations with the largest weight in the total wave function. Hence, the algorithm can be understood as a sophisticated way of selecting the most important configurations and is therefore capable of approximating wave functions of arbitrary degree of multi-configurational character given that the number of renormalized block states required to converge the calculation is low enough to render the calculation affordable. Importantly, such situations are detectable within vDMRG itself through well-established truncation errors.

We note that in Refs.~\citenum{Christiansen2004_SecondQuantization} and \citenum{Wang2009_SQMCTDH} a more general second-quantized form of the vibrational Hamiltonian was proposed, where a general one-dimensional basis set is applied for each mode, the eigenfunctions of the harmonic-oscillator Hamiltonian 
being only one specific choice.
Although these more general formulations can, in principle, be exploited within vDMRG, they would increase its computational cost significantly. 
The reason is that, when a general basis set is employed for each mode, excitations of a mode are represented by different basis functions (modals) with one pair of creation and annihilation operators to be introduced for each modal, and then the total number of local second-quantized elementary operators scales as $L × N_\text{max}$, where $N_\text{max}$ is the maximum number of modals for each mode.
However, when Hermite polynomials are employed as local basis set (as chosen for this work), the number of local elementary operators grows as $L$, because one pair of creation and annihilation operators is needed for each mode only.
The more general basis functions are usually accompanied by the $n$-mode representation of the potential (see e.g., Refs.~\citenum{Bowman2003_multimode} and \citenum{Gerber1995_vscf}, which is more flexible than the Taylor expansion in Eq. (4).
The total number of terms in the Hamiltonian will be approximately the same for both representations. However, the number of sites on the vDMRG lattice will equal the number of modals in the $n$-mode representation with a general basis, while it equals the number of modes in the Taylor expansion of the potential with harmonic oscillator basis functions.
The vDMRG algorithm and our implementation allow one to work with any type of basis functions and potential. 
Here, we focus on the Taylor expansion of the potential with a harmonic oscillator basis in this initial implementation and extend this to a more general formulation in future work, as this extension would be beyond the scope of this introduction of the vDMRG algorithm.
To conclude, we note that the $n$-mode representation of the potential would in principle allow one to develop a state-specific version of vDMRG, where different basis functions are used for each mode. Therefore it is possible to target directly highly-excited vibrational states when combined with an advanced diagonalization algorithm (\textit{vide infra}), where a specific energy range can be selected.

\section{Implementation and Methodology}

With the second-quantized form of the Watson Hamiltonian, Eq.~(\ref{eq:WatsonHamiltonian}), the implementation of the MPO-based DMRG algorithm presented in Ref.~\citenum{Keller2015} can be extended to the calculation of vibrational properties. 
We implemented this Hamiltonian into our \textsc{QCMaquis-V} program\cite{Keller2015,knec16} that is based on an MPS library developed
for spin Hamiltonians\cite{bau11,Dolfi2014_ALPSImplementation} and extends the \textsc{QCMaquis} program to the calculation of vibrational energies.
\textsc{QCMaquis} provides an implemention of the full electronic 
Hamiltonian with up to four fermionic elementary creation and annihilitation operators. For the generalization of this implementation,
we needed to 1) implement the commutation symmetry of bosonic creation and annihilation operators (to replace the anticommutation symmetry
of the fermionic creation and annihiliation operators expressed in terms of $\widehat F_s$ operators in Ref.~\citenum{Keller2015}),
2) allow for arbitrary occupations $\sigma_i$, and implement MPOs for sequences of more than four elementary creation and annihilation
operators. All of this was comparatively easy to accommodate by virtue of the modular and object-oriented structure of \textsc{QCMaquis-V}. 

The input data for our vDMRG calculations (i.e., harmonic frequencies, anharmonic force constants, Coriolis coupling constants) were
either taken from the literature data or, for the dipeptide {SarGly$^+$}, calculated with the \textsc{Gaussian} program.\cite{g16.a03}
Detailed information on the electronic structure methods applied for the generation of the PESs is specified in the respective sections below.

We emphasize that the variational optimization of the MPS provides the anharmonic zero-point vibrational energy (ZPVE). 
To also obtain transition frequencies, the energy of vibrationally excited states must be determined.
This is accomplished by an optimization in the space orthogonal to the one spanned by the lower-energy vibrational states through an appropriate projection operation as described in Ref.~\citenum{Keller2015}.
Transition energies $h\nu_k$ are then calculated as
\begin{align}
h\nu_k = E_{k}-\text{ZPVE} \, .
\label{trans_energy}
\end{align}


\section{Results}
\subsection{Three-atomic molecule: ClO$_2$}

To validate our implementation of vDMRG, we calculated the vibrational states of the triatomic molecule ClO$_2$ in its electronic ground state 
of X$^2B_1$ symmetry because an accurate PES obtained from multi-reference configuration interaction (MRCI) calculations is available in the literature.\cite{Peterson1998_Variational,Crittenden2015_PyPES}
Fully converged vibrational energies up to 3300~cm$^{-1}$ were calculated from this PES with variational approaches.\cite{Peterson1998_Variational,Crittenden2015_PyPES}
The high accuracy of these calculated frequencies was demonstrated by comparison to high-resolution experimental data.\cite{Ortigoso1991_IRCl2OFirstBand,Ortigoso1992_IRCl2OOtherBands,Ortigoso1993_IRCl2OCombBands} 
These theoretical results\cite{Peterson1998_Variational} will be our reference in the following to assess the convergence of our vDMRG calculations.

Before discussing the results of our calculations, we compare the vDMRG approach with the variational approaches of the reference papers.
In the study by Peterson,\cite{Peterson1998_Variational} the PES is expressed as a Taylor expansion in terms of the internal coordinates (the two bond lengths and the bond angle) of the molecule.
The variational calculations of the vibrational problem are then carried out directly in these coordinates, with Morse oscillator eigenfunctions as a basis set for the two stretching coordinates and a DVR for the bond angle.\cite{Bacic1989_DVR} 
In a more recent work,\cite{Crittenden2015_PyPES} the quartic force field in internal coordinates is first transformed to a sixth-order force field in Cartesian normal coordinates, by a direct tensor transformation.\cite{Hoy1972_Tensor,Crittenden2015_PyPES}
The variational calculation is then carried out in a basis of harmonic-oscillator wave functions.
In this work, we employ the latter representation of the PES in Cartesian normal coordinates, because the second-quantized expression of the vibrational Hamiltonian given in Eq.~(\ref{eq:WatsonHamiltonian}) 
holds for a basis of  harmonic-oscillator wave functions.
The two representations are, in principle, different because internal coordinates are non-linear functions of the Cartesian normal modes, but they result in equal fundamental frequencies\cite{Peterson1998_Variational,Crittenden2016_PyVCI} and are therefore equivalent close to the equilibrium structure. 
Although more recent results are reported in Ref.~\citenum{Crittenden2016_PyVCI}, the data reported in Ref.~\citenum{Peterson1998_Variational} will serve as our reference, because the authors of Ref.~\citenum{Peterson1998_Variational} also included overtones and combination bands in their discussion.

\begin{table*}[htb!]
  \begin{center}
    \caption{vDMRG zero-point energy and energies of the 17 lowest vibrational transitions (in cm$^{-1}$) calculated from Eq.~(\ref{trans_energy}) for ClO$_2$.  Each state was calculated with different values for the parameters $N_{\text{max}}$ and $m$.
  \label{tab:Cl2O} 
}
    \begin{tabular}{cc|ccc|cc|c}
	\hline \hline
	  \multirow{2}{*}{Assign.}  &  \multirow{2}{*}{Ref.} & \multicolumn{3}{c|}{$N_{\text{max}} = 10$} & \multicolumn{2}{c|}{$N_{\text{max}} = 12$} & \multirow{2}{*}{Harm.} \\
	                      &             &   $m=2$    &    $m=5$    &     $m=10$  &      $m=5$        &       $m=10$        &            \\
	  \hline
	   ZPE                &             &   1264.5   &    1264.5   &    1264.5   &     1264.5        &       1264.5        &            \\
	   $\nu_2$            &    449.9    &    449.5   &     449.5   &     449.5   &      449.5        &        449.5        &   455.62   \\
	  $2\nu_2$            &    898.7    &    898.0   &     898.1   &     898.1   &      898.0        &        898.2        &   911.24   \\
	   $\nu_1$            &    940.7    &    940.7   &     940.7   &     940.7   &      940.7        &        940.5        &   960.15   \\
	   $\nu_3$            &   1105.5    &   1105.2   &    1105.2   &    1105.2   &     1105.2        &       1105.2        &  1127.82   \\
	  $3\nu_2$            &   1346.5    &   1345.6   &    1345.5   &    1345.5   &     1345.5        &       1345.8        &  1366.86   \\
	  $\nu_1+\nu_2$       &   1386.5    &   1386.2   &    1386.2   &    1386.3   &     1386.1        &       1385.8        &  1415.77   \\
	  $\nu_2+\nu_3$       &   1549.9    &   1548.7   &    1548.7   &    1548.7   &     1548.6        &       1548.6        &  1573.44   \\
	  $\nu_1+2\nu_2$      &   1831.2    &   1830.7   &    1830.6   &    1830.6   &     1830.6        &       1828.8        &  1871.39   \\
	  $2\nu_1$            &   1872.1    &   1872.4   &    1872.3   &    1872.3   &     1872.3        &       1872.2        &  1920.30   \\
	  $2\nu_2+\nu_3$      &   1993.2    &   1991.2   &    1991.2   &    1992.8   &     1991.1        &       1991.1        &  2039.06   \\
	  $\nu_1+\nu_3$       &   2029.3    &   2029.3   &    2028.9   &    2029.1   &     2029.2        &       2029.2        &  2087.97   \\
	  $2\nu_3$            &   2200.1    &   2201.0   &    2199.5   &    2199.5   &     2199.5        &       2199.6        &  2255.64   \\
	  $2\nu_1+\nu_2$      &   2313.8    &   2318.4   &    2313.5   &    2313.6   &     2313.6        &       2313.7        &  2375.92   \\
	  $\nu_1+\nu_2+\nu_3$ &   2469.8    &   2468.3   &    2468.5   &    2468.5   &     2468.1        &       2468.4        &  2543.59   \\
	  $3\nu_1$            &   2794.0    &   2797.5   &    2794.6   &    2794.6   &     2794.7        &       2794.7        &  2880.45   \\
	  $2\nu_1+\nu_3$      &   2943.2    &   2947.9   &    2943.8   &    2943.2   &     2943.5        &       2943.7        &  3048.12   \\
	  \hline \hline                                                       
    \end{tabular}                                                   
  \end{center}
  \end{table*}     

As has already been pointed out above, two parameters control the convergence of the vDMRG vibrational wave function and energy.
The first, $N_\text{max}$, corresponds to the number of basis functions that describe each mode and its value must be increased with increasing anharmonicity of a given mode.
The second parameter, the number of renormalized block states $m$, controls the degree of dimension reduction of the final MPS wave function and has to be increased with increasing anharmonic mode coupling.
These two parameters are, in general, independent and the convergence with respect to both has to be analyzed. 

vDMRG vibrational energies up to 3000~cm$^{-1}$ calculated with different values for $\NMax$ and $m$ are reported in Table~\ref{tab:Cl2O}, together with the harmonic results and theoretical reference data taken from Ref.~\citenum{Peterson1998_Variational}
The number of vDMRG sweeps was set to 40 for all states and convergence with respect to the number of sweeps was found in all cases.
With ten basis functions per mode ($\NMax=10$), converged results are obtained for energies up to 2000~cm$^{-1}$ with as little as two renormalized states, and variations below 0.1 cm$^{-1}$ are obtained by increasing $m$ to 5 and 10. 

The vDMRG results are in good agreement with the theoretical reference values and confirm that a sixth-order Taylor expansion in Cartesian normal modes is sufficient to obtain accurate fundamental frequencies.
However, for some of the higher-energy transitions, two renormalized states are not sufficient to reach convergence of energies below 1 cm$^{-1}$. 
This is particularly evident for the $2\nu_1+\nu_2$, $3\nu_1$ and $2\nu_1+\nu_3$ transitions, where results obtained with $m=2$ and $m=10$ deviate by more than 2 cm$^{-1}$. 
It should be noted that these are coupled vibrational modes with the highest energy among those considered in this table, and therefore show the largest anharmonic corrections.
For this reason, a higher number of renormalized states is required to recover accurate variational energies. 
By further increasing the dimension of the local basis to twelve, no further variations occur and the energies can therefore be considered
converged with respect to the basis set size.
In Table~S1 of the Supporting Information, the results obtained with a quartic potential are reported.

To conclude, the results obtained for ClO$_2$ show the robustness of our vDMRG implementation.

\subsection{Parameter dependence: CH$_3$CN}

To investigate the convergence of the vDMRG energies and wave functions with respect to $m$ and $N_\text{max}$ in more detail, we calculated the 33 lowest vibrational states of acetonitrile in its electronic ground state for various choices of these parameters.
A quartic force field for acetonitrile was reported in Ref.~\citenum{begu05} and has become a classical benchmark for newly developed variational approaches.\cite{avil11,lecl14,lecl16,lecl16a,Rakhuba2016_TensorTrain}
We derived the full quartic force field with the symmetry relations given in Refs.~\citenum{henr61,henr65,picc15}.
From a footnote in Ref.~\citenum{lecl14}, which refers to a private communication, it appears that there is apparently some confusion in the literature about the application of these symmetry relations.
Therefore, we included the explicit expressions for the symmetry relations that we applied in the Supporting Information in order to provide unambiguous information about the underlying force field of our vDMRG calculations.
Note that in Ref.~\citenum{begu05} only force constants larger than 7~cm$^{-1}$ are reported, and therefore minor deviations of our data
from the theoretical reference data in Ref.\ \onlinecite{avil11} are to be expected.

\begin{figure}[htb!]
\includegraphics[width=0.5\textwidth]{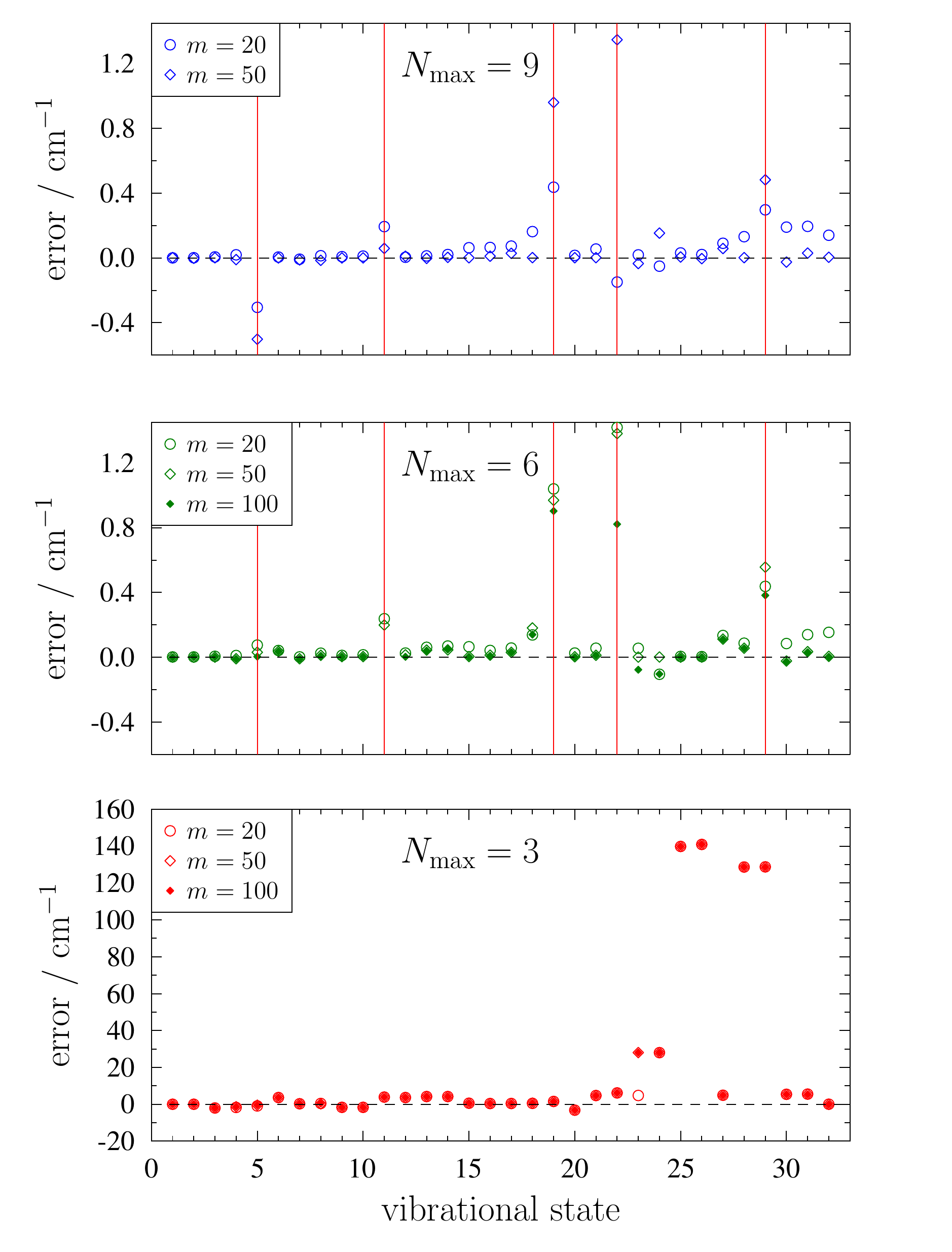}
\caption{Relative error of the vibrational energies (in cm$^{-1}$) for $N_\mathrm{max}$ = 3 (lower panel), 6 (middle panel), and 9 (upper panel) and $m$ = 20, 50, and 100. All errors are with respect to the largest parameters ($N_\mathrm{max}= 9$ and $m=100$). Red vertical lines indicate states with errors larger than 0.2~cm$^{-1}$ for at least one of the parameter sets.
\label{MeCN_figure}
}
\end{figure}

Table~\ref{MeCN_table} contains the vDMRG results for $m=20,50,100$ and $N_\text{max} = 3,6,9$ for the 11 lowest vibrational states along with their deviation from the results of Ref.~\citenum{avil11}.
We applied 50~DMRG sweeps for each state and ensured that the energy was converged with respect to the number of sweeps.
If that was not the case, we had to include more sweeps until the energy was converged to less than 0.001~cm$^{-1}$.
The assignment and the symmetry were identified by comparison of the energies to the results of Ref.~\citenum{avil11}. Since the coefficients of the configurations are not as directly available for the MPS wave function as they are in case of standard VCI calculations, the description of the character and symmetry of a given state would require an additional computational step. It has been demonstrated for the electronic structure problem (Refs.\citenum{Moritz2007_SlaterDecomposition} and \citenum{bogus11}) that a sampling of the most important configurations yields an approximate expression of the wave function in determinant basis. A similar strategy will be exploited in future work to facilitate the determination of the character and symmetry of the vibrational vDMRG states.
Apparently, the small value of $m=20$ is sufficient to guarantee convergence to less than 0.1~cm$^{-1}$ for the states considered in Table~\ref{MeCN_table} when a local basis size of at least $N_\text{max}=6$ is employed.
This is encouraging because the calculation of a single vibrational state with $m=20$ takes only around 15~\% of the time of the same calculation with $m=100$.
Since this factor is approximately the same for all states, the savings are huge when a large number of vibrational states is to be calculated, which is 
obviously the case for large molecules.
Figure~\ref{MeCN_figure} shows the error of a given set of parameters with respect to the largest calculation with $m=100$ and $N_\text{max}=9$.
Clearly, if the local basis size is too small, the errors can become very large ($>100$~cm$^{-1}$), especially for higher excited states because the harmonic wave functions for the overtones are simply not included in the basis set.
The deviations increase for higher excited states since each excited state is optimized in the space orthogonal to all lower-lying states.
Of course, the optimization will become unreliable if the wave functions of the lower-lying states are not converged.
However, for both, $N_\text{max}=6$ and 9, the deviations do not exceed 1.3~cm$^{-1}$ and are most often below 0.2~cm$^{-1}$.
In the light of these results and those obtained in the previous section, we conclude that a local basis set size of at least $N_\text{max} = 6$ is required, whereas the number of renormalized block states $m$ can be as small as 10 or 20.

Some vibrational states are certainly more sensitive to $m$ (red vertical lines in Figure~\ref{MeCN_figure}) than others.
In the first four cases (states 5, 11, 19, and 22), it is the $A_{1/2}$ component of a vibrational state that has almost the same energy as its corresponding degenerate component of $E$ symmetry.
The fifth state with a discrepancy of up to 0.5~cm$^{-1}$ is, however, one component of an $E$ state (corresponding to 4~$\nu_{11}$).
We conclude that special care must be taken to fully converge states that are close in energy or even degenerate if an accuracy below one reciprocal
centimeter is desired. In regions with a high density of states (such as the C-H stretch region in organic molecules) this becomes especially challenging. The convergence with respect to all DMRG specific parameters has to be monitored such that each individual state is fully converged. If this is not ensured, the calculation of higher excited states can become inaccurate because these states are optimized in the space orthogonal to only partially converged lower states and errors might accumulate. This limitation can be in principle overcome with modified diagonalization algorithms. Among them, Lanczos (Refs.~\citenum{coriani12} and \citenum{coriani13}) and modified Davidson (see Refs. \citenum{chan07,li11,petrenko17} for example of application both to electronic and vibrational structure problems) algorithms are particularly appealing in this respect. Additionally, subspace diagonalization algorithms, that have been applied both to electronic structure calculations\cite{neugebauer11} and to vibrational problems,\cite{neugebauer03} would improve the efficiency of vDMRG, and work is in progress in this respect.

Deviations with respect to the reference values on the order of 1~cm$^{-1}$ remain even for our most accurate calculations (see Table~\ref{MeCN_table}).
Since our calculations are converged with respect to the number of sweeps and size of the local basis set, the discrepancy could be 
a result of the symmetry relations applied to construct the full quartic force field from the force constants of Ref.~\citenum{begu05}.

\begin{table*}[htb!]
\begin{center}
\caption{Zero-point energies and ten lowest transition energies (in cm$^{-1}$) calculated from Eq.~(\ref{trans_energy}) for CH$_3$CN obtained with the quartic force-field of Ref.~\citenum{begu05}. Deviations with respect to the reference values from Ref.~\citenum{avil11} are given in parentheses. The vDMRG vibrational energies were calculated with different values for the parameters $N_{\mathbf{max}}$ and $m$.
\label{MeCN_table}
}
\begin{tabular}{ccr|rrr|rrr|rrr}
\hline \hline
\multicolumn{2}{c}{}     & reference             & \multicolumn{3}{l|}{$N_{\mathbf{max}} = 3$} & \multicolumn{3}{l|}{$N_{\mathbf{max}} = 6$} & \multicolumn{3}{l}{$N_{\mathbf{max}} = 9$} \\
Assign. & Sym. & energy &$m=20 $&$m=50$ &$m=100$ &$m=20 $&$m=50$ &$m=100$ &$m=20 $&$m=50$ &$m=100$ \\
\hline
ZPE            &          & 9837.498 & 9837.590 & 9837.589 &9837.589 &9836.570  & 9836.568  & 9836.568 & 9836.566  & 9836.565 &9836.565  \\
$\nu_{11}$  & $E$   &361.08      & 361.03     & 361.03     & 361.03   & 361.00      & 361.00      & 361.00    & 361.00       & 361.00     &361.00      \\	 
                   &         &                  & (-0.05)     & (-0.05)      & (-0.05)   & (-0.08)       & (-0.08)     & (-0.08)     &(-0.08)         &(-0.08)      &(-0.08)      \\ 
                    &         &361.15       & 361.04     & 361.03     & 361.03   & 361.01      & 361.01     & 361.01     &361.01       & 361.00     & 361.01     \\	 
                    &         &                  & (-0.11)      & (-0.12)     & (-0.12)    & (-0.07)      & (-0.07)      & (-0.07)     &(-0.07)       &(-0.08)       &(-0.07)      \\
$2\nu_{11}$ & $E$  &  723.25    & 721.22      & 721.21     & 721.21   & 723.22      & 723.21     & 723.21     &723.22      & 723.21     & 723.21     \\	 
                   &          &                 & (-2.03)      & (-2.04)      & (-2.04)   & (-0.03)       & (-0.04)      & (-0.04)     &(-0.03)       &(-0.04)     &(-0.04)     \\
                    &          & 723.63     & 721.53     & 721.90      & 721.90   & 723.24      & 723.22     & 723.22    &723.25       & 723.22     & 723.23     \\	 
                    &          &                 & (-2.10)      & (-1.73)      & (-1.73)   & (-0.39)        & (-0.41)     & (-0.41)    &(-0.38)       &(-0.41)      &(-0.40)     \\ 
$2\nu_{11}$ &$A_1$&724.35     & 722.87      & 723.27      & 723.27  & 723.82       & 723.78    & 723.75     &723.44      & 723.25      & 723.75    \\ 
                   &           &                & (-1.48)      & (-1.08)      & (-1.08)   & (-0.53)       & (-0.57)     & (-0.60)     &(-0.91)       &(-1.10)       &(-0.6)       \\	 
$\nu_4$       &$A_1$ & 900.78   & 904.35      & 904.34      & 904.34   & 900.72      & 900.71     & 900.71    &900.68      & 900.67      & 900.67     \\	 
                    &            &               & (3.57)       & (3.56)        & (3.56)    & (-0.06)       & (-0.07)     & (-0.07)     &(-0.10)      &(-0.11)       &(-0.11)       \\ 
$\nu_9$       &$E$     &1034.40   & 1033.55   & 1033.55     & 1033.55 & 1033.27   & 1033.26   & 1033.26   &1033.26   & 1033.26    &1033.27    \\	 
                   &            &                & (-0.85)     & (-0.85)       & (-0.85)   & (-1.13)      & (-1.14)      & (-1.14)     &(-1.14)     &(-1.14)       &(-1.13)     \\	 
                    &            & 1034.74  & 1033.74   & 1033.55     & 1033.74 & 1033.31  & 1033.30    & 1033.29   &1033.30   & 1033.27   & 1033.29    \\	 
                   &            &                & (-1.00)     & (-1.19)       & (-1.00)    & (-1.43)      & (-1.44)     & (-1.45)      &(-1.44)     &(-1.47)       &(-1.45)     \\	 
$3\nu_{11}$ &$A_1$ & 1087.27  & 1084.97   & 1084.97     & 1084.97 & 1086.63   & 1086.62   & 1086.62    &1086.62   & 1086.62   & 1086.62   \\ 
                    &            &                & (-2.30)     & (-2.30)        & (-2.30)   & (-0.64)      & (-0.65)     & (-0.65)      &(-0.65)     &(-0.65)       &(-0.65)     \\	 
$3\nu_{11}$ &$A_2$ & 1087.40  & 1084.98   & 1084.97      & 1084.97 & 1086.63   & 1086.62  & 1086.62    &1086.63   & 1086.62   & 1086.62   \\	 
                     &           &                & (-2.42)     & (-2.43)        & (-2.43)     & (-0.77)     & (-0.78)    & (-0.78)      &(0.77)       &(-0.78)      &(-0.78)     \\
\hline \hline
\end{tabular}
\end{center}
\end{table*}

\subsection{Higher-order expansion of the potential energy surface: Ethylene}

The third example selected to demonstrate the reliability of vDMRG is ethylene ({C$_2$H$_4$}). 
Here, we adapt a highly accurate \textit{ab initio} PES from the literature\cite{Delahaye2014_EthylenePES} and compare the results with previous calculations and experimental data.\cite{Georges1999_C2H4Exp}
Two decades ago, an accurate PES for ethylene was constructed from CCSD(T) calculations by Martin and coworkers.\cite{Martin1995_C2H4Potential} 
With this PES, vibrational energies and spectra were determined with various variational approaches.\cite{Avila2011_C2H4PrunedBasis,Carter2012_C2H4Multimode} 
Recently, a new PES was constructed from CCSD(T) calculations with a large quadruple-$\zeta$ atomic orbital basis, from which more accurate vibrational properties could be calculated.\cite{Delahaye2014_EthylenePES} 
We applied the latter PES in our vDMRG calculations.
The analytical expression in terms of internal coordinates can be found in Ref.~\citenum{Delahaye2014_EthylenePES}.
Since our implementation of vDMRG is based on Cartesian normal coordinates as a reference coordinate system, we converted the PES to a Taylor series expansion in terms of these coordinates with the procedure discussed in Refs.~\citenum{Crittenden2015_PyPES} and \citenum{Crittenden2016_PyVCI} .
Fourth- and sixth-order Taylor series expansion of the potential, with and without the inclusion of Coriolis effects, were considered in the vDMRG calculations with 20 sweeps for each state.
The order of the Taylor expansion was chosen based on the variational results reported in Ref.~\citenum{Crittenden2016_PyVCI}, where it was shown that a sixth-order expansion is sufficient to reach convergence within 1 cm$^{-1}$. 
In the theoretical reference work,\cite{Delahaye2014_EthylenePES} Coriolis terms were included only for two transitions, and therefore they have been neglected also in our vDMRG calculations. 

\begin{figure}[htb!]
	\centering
	\includegraphics[width=.5\textwidth]{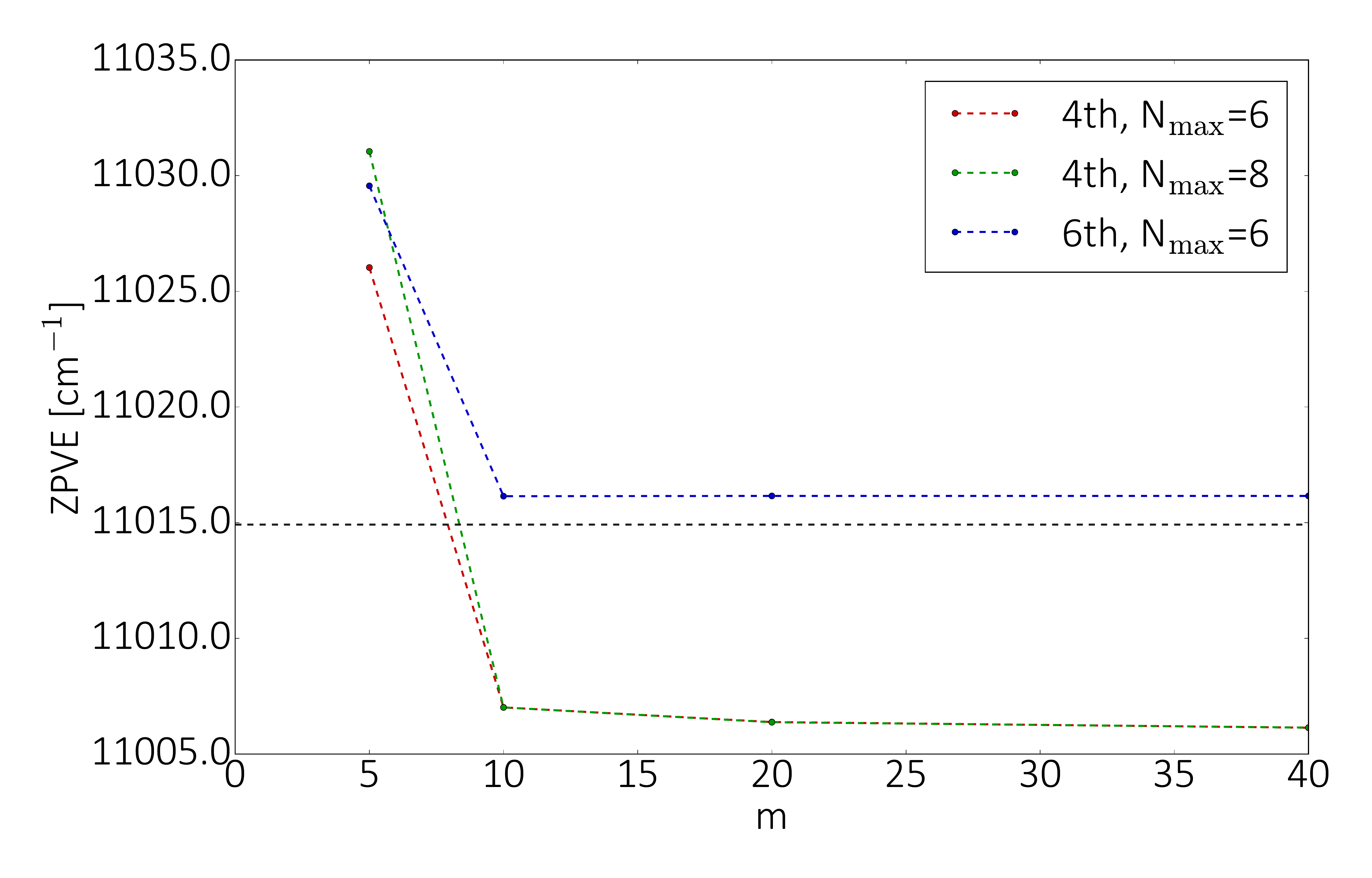}
	\caption{vDMRG ZPVE of {C$_2$H$_4$} as a function of the number of renormalized block states $m$. The reference ZPVE from
		     Ref.~\citenum{Delahaye2014_EthylenePES} is represented as a black dotted line.
	\label{fig:C2H4_conv_m}
}
\end{figure}

Vibrational energies for the ground and the first two excited states of {C$_2$H$_4$} are reported in Tables S2, S3, and S4 of the Supporting Information.
The data reported in Table S2 highlight the effect of each of the two parameters $m$ and $N_\text{max}$ on the ZPVE. 
First, we discuss the dependence of the ZPVE on $m$ for a given order of the Taylor expansion of the potential (a fixed Hamiltonian) and then for a given value of $N_\text{max}$ (a fixed local basis set).
The data are also shown in Figure~\ref{fig:C2H4_conv_m}.
With five renormalized states, inaccurate results with deviations larger than 10~cm$^{-1}$ with respect to the converged values, are obtained. 
Convergence (with deviations $<$ 1~cm$^{-1}$) is reached with ten renormalized block states for all Hamiltonians (i.e., with 
a quartic and a sextic force field).
This indicates that the further inclusion of high-order terms in the potential will not affect significantly the structure of the Hamiltonian, i.e., it will not introduce further long-range correlations that would require a larger number of renormalized block states and DMRG sweeps to converge the wave function.

Regarding the convergence of the ZPVE with respect to the size of the local basis $\NMax$, fully converged results are obtained for $\NMax=6$, and with higher values (8 and 10), the change in the ZPVE is below 0.01 cm$^{-1}$. 
For {C$_2$H$_4$}, neither highly anharmonic, large-amplitude vibrations are present nor resonances that usually cause strong variational corrections. 
Therefore, a reliable representation for the vibrational wave function can be obtained from a relatively small number of harmonic-oscillator eigenfunctions as local basis functions.

As noted above for {ClO$_2$}, a virtually infinite-order Taylor expansion in Cartesian normal coordinates would be required to obtain a potential that is equivalent to the one reported in Ref~\citenum{Delahaye2014_EthylenePES}, and therefore convergence with respect to the order of the Taylor series expansion of the potential must be studied.
With a fourth-order PES, the ZPVE is underestimated by $\approx 7$~cm$^{-1}$ for all choices of $m$ and $N_\text{max}$, and the further inclusion of fifth- and sixth-order terms reduces the error to less than 1~cm$^{-1}$, therefore providing satisfactory convergence. 
This result is in line with the analysis reported in Refs.~\citenum{Crittenden2015_PyPES} and \citenum{Crittenden2016_PyVCI}, where results converged within 1 cm$^{-1}$ were obtained from a full sextic potential without including Coriolis couplings.
The vDMRG ZPVE obtained with the sextic potential and with $\NMax=6$ and $m=20$ is 11016.15~cm$^{-1}$, and agrees well with the theoretical reference value of 11014.91~cm$^{-1}$,\cite{Delahaye2014_EthylenePES} which was calculated without Coriolis couplings.

In Tables S3 and S4, a similar analysis is carried out for the first two fundamental bands of ethylene. 
Again, converged results were obtained with $\NMax=6$, and only minor changes are observed with $\NMax=8$. 
Once again, the convergence with respect to the number of renormalized block states $m$ is slower than for the ZPVE and discrepancies of more than 1~cm$^{-1}$ can be observed for $m=10$. 

\newcommand{\Mlt}[1]{\multicolumn{2}{c}{#1}}
\newcommand{\Mltc}[1]{\multicolumn{2}{c|}{#1}}
\begin{table*}[htb!]
  \begin{center}
    \caption{Comparison of the 18 lowest vibrational transition energies calculated from Eq.~(\ref{trans_energy}) and zero-point energies (in cm$^{-1}$) of {C$_2$H$_4$} calculated with vDMRG applying a fourth- (quartic) and a sixth-order (sextic) Taylor expansion of the PES along with theoretical reference data from Ref.~\citenum{Delahaye2014_EthylenePES}, and
             experimental data from Ref.~\citenum{Georges1999_C2H4Exp}. The vDMRG vibrational energies were calculated with different values for the parameters $N_{\mathbf{max}}$ and $m$.
	\label{tab:C2H4_Energies}
}
\begin{tabular}{cC|ccccc|cc}
      \hline\hline
         \multirow{2}{*}{Mode} & \multirow{2}{*}{Assign.} & \Mlt{Quartic} & \Mlt{Sextic} & Sextic + C & \\
         &  &  \NMax=6  & \NMax=6 &  \NMax=6 & \NMax=6 &  \NMax=6  & & \\
         &  &  $m$=10     &  $m$=20   &  $m$=10    & $m$=20    &  $m$=10     & Ref.\cite{Delahaye2014_EthylenePES} & Exp.\cite{Georges1999_C2H4Exp} \\[7pt]
      \hline
       ZPVE  &                &   11007.01      &    11006.19     &    11016.14    &   11016.15    & 11021.39  &  11014.91   &           \\       
          1  & \nu_{10}       &     810.27      &      809.03     &      831.13    &     831.17    &   834.64  &    822.42   &   825.92  \\
          2  & \nu_8          &     916.61      &      915.29     &      933.48    &     933.47    &   942.26  &    934.29   &   939.86  \\
          3  & \nu_7          &     929.76      &      928.31     &      948.26    &     948.26    &   957.15  &    949.51   &   948.77  \\
          4  & \nu_4          &    1008.46      &     1007.13     &     1018.26    &    1018.26    &  1026.64  &   1024.94   &  1025.58  \\
          5  & \nu_6          &    1217.88      &     1217.17     &     1227.08    &    1227.05    &  1229.68  &   1224.96   &  1225.41  \\
          6  & \nu_3          &    1339.55      &     1338.87     &     1343.46    &    1343.46    &  1344.25  &   1342.96   &  1343.31  \\
          7  & \nu_{12}       &    1431.74      &     1430.47     &     1441.50    &    1441.52    &  1444.94  &   1441.11   &  1442.44  \\
          8  & \nu_2          &    1619.69      &     1622.11     &     1628.26    &    1629.04    &  1628.41  &   1624.43   &  1626.17  \\
          9  & 2\nu_{10}      &    1626.57      &     1625.56     &     1682.30    &    1682.18    &  1685.54  &   1658.39   &  1664.16  \\
         10  & \nu_8+\nu_{10} &    1723.42      &     1722.77     &     1769.89    &    1770.02    &  1764.78  &   1757.70   &  1765.78  \\
         11  & \nu_7+\nu_{10} &    1739.03      &     1729.53     &     1787.02    &    1786.99    &  1781.17  &   1778.34   &  1781.01  \\
         12  & \nu_4+\nu_{10} &    1814.03      &     1810.47     &     1840.52    &    1852.60    &  1865.89  &   1848.61   &  1851.51  \\
         13  & 2\nu_8         &    1830.01      &     1826.01     &     1878.67    &    1878.42    &  1890.40  &   1873.73   &  1879.72  \\
         14  & \nu_7+\nu_8    &    1830.71      &     1827.96     &     1886.15    &    1886.16    &  1897.34  &   1885.12   &  1888.63  \\
         15  & 2\nu_7         &    1855.82      &     1852.23     &     1905.44    &    1906.02    &  1901.04  &   1901.61   &  1899.74  \\
      \hline\hline
    \end{tabular}
  \end{center}
\end{table*}

A complete list of the 18 lowest vibrational frequencies (including both overtones and combination bands) of {C$_2$H$_4$} calculated with vDMRG is collected in Table~\ref{tab:C2H4_Energies}. In view of these results, all calculations were carried out with $\NMax = 6$, with both a quartic and a sextic potential and 
with 10 and 20 renormalized block states $m$.
These results confirm the trend already found for the three lowest-energy states. 
First of all, the comparison of the results shows that ten renormalized states are sufficient to achieve an accuracy of 1~cm$^{-1}$ for the lowest energy states.
However, the difference between the results obtained with $m$=10 and $m$=20 increases for higher excited states but remains below 10~cm$^{-1}$, even for states up to 1800~cm$^{-1}$ above the ZPVE.
In analogy with acetonitrile, the largest deviations are usually obtained for states involved in resonances, as for example the $2\nu_7$ and $\nu_7+\nu_8$ states in Fermi resonance.
It has already been widely discussed in the literature that variational corrections are particularly relevant for resonant states.\cite{Bloino2012_GVPT2,Krasnoshchekov2014_VVPT2,Rosnik2014_VPT2K}
For this reason, a higher number of renormalized block states $m$ is required to obtain converged energies. 

These observations hold for both the quartic and the sextic potential.
Moreover, for the latter potential, a slower convergence with respect to $m$ is observed for vibrational states with higher energy. 
The frequencies we calculate from the sextic potential agree significantly better with the theoretical reference data than the frequencies we obtained from the quartic potential, especially for higher frequencies, above 1500~cm$^{-1}$. 

Finally, we calculated vibrational wave functions and energies from a sextic potential with Coriolis coupling.
Rotational effects were also considered in the theoretical reference work\cite{Delahaye2014_EthylenePES} for two, high-energy fundamental transitions ($\nu_{11}$ and $\nu_9$, with harmonic frequencies of 3140.91~cm$^{-1}$ and 3248.71~cm$^{-1}$, respectively) by fitting the energy of more than 50 calculated rovibrational states. 
For all other transitions, rotational effects were neglected. 
The rotational correction amounted to approximately 3~cm$^{-1}$, which is about the same order of magnitude as the discrepancy between vDMRG and experimental data. 
The results obtained including Coriolis terms for the rotational ground state (reported in Table~\ref{tab:C2H4_Energies} as "Sextic + C") confirm that rotational corrections are relevant for several bands (e.g., $\nu_7$ and $\nu_8$), for which this effect amounts to up to 9 cm$^{-1}$. 
In general, the inclusion of rotational corrections improves the agreement between vDMRG and experimental data. 
It should be noted that, in some cases (as, e.g., for the $\nu_4$ fundamental band), the discrepancy between vDMRG and theoretical 
reference results without rotational correction is significant (above 5~cm$^{-1}$). 
The inclusion of a rotational correction however enhances the agreement with the experimental data and indicates a better agreement of the variational calculations than the theoretical reference work.\cite{Delahaye2014_EthylenePES}

\subsection{Application to a large molecule: The {SarGly$^+$} dipeptide}

\begin{figure}[htb!]
	\centering
	\includegraphics[width=.49\textwidth]{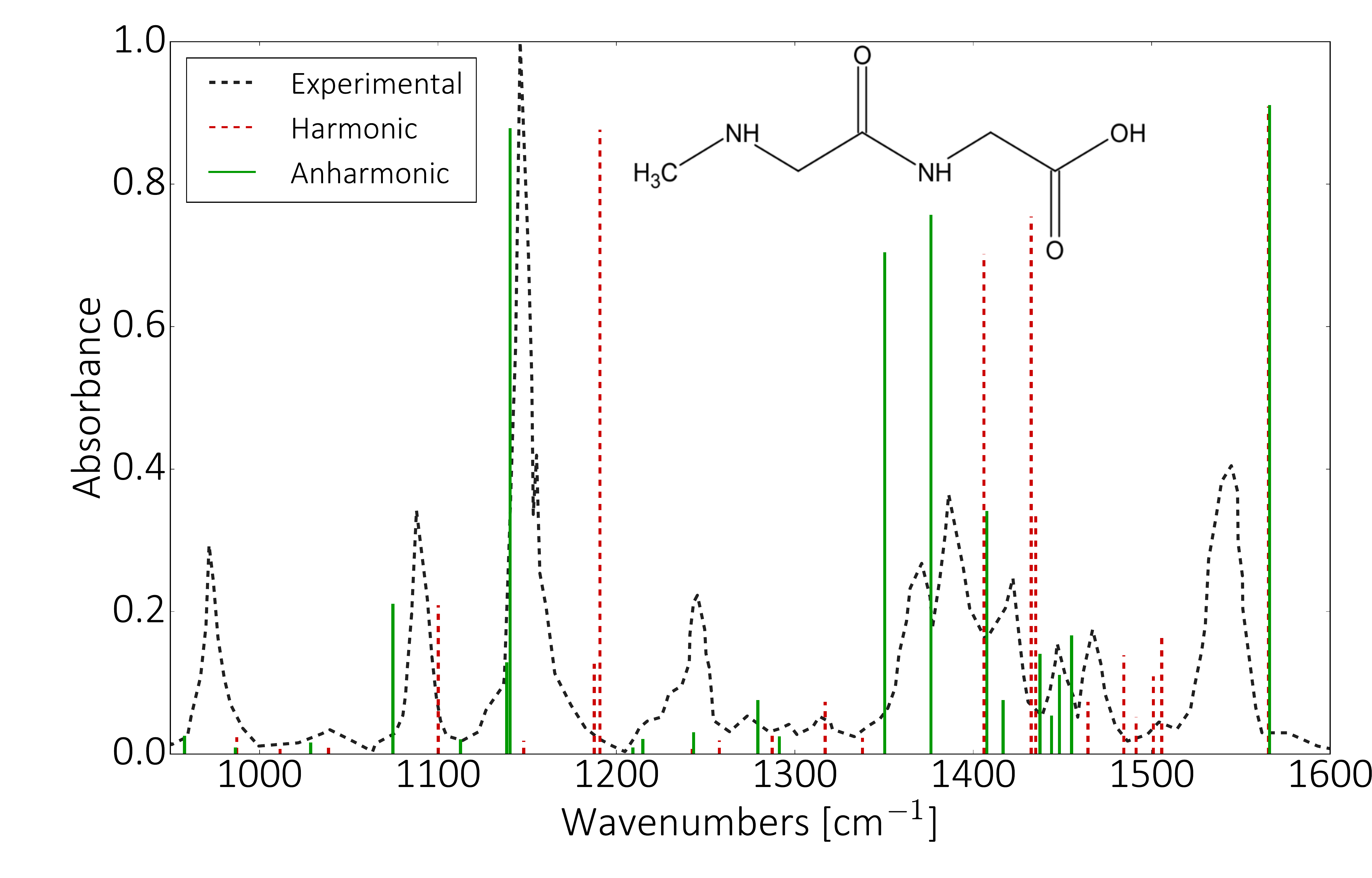}
	\caption{Experimental\cite{Johnson2014_SarGlyExp} and theoretical infrared spectrum of {SarGly$^+$} computed from harmonic (red lines) and anharmonic 
		     vDMRG frequencies (green lines). The parameters of the vDMRG calculations are $\NMax=6$ and $m=20$. In all cases, harmonic intensities were calculated.
	\label{fig:IR_dipeptide}
}
\end{figure}

Our last example is the protonated sarcosine-glycine dipeptide (referred to as {SarGly$^+$} in the following, whose
Lewis structure is shown in Figure \ref{fig:IR_dipeptide}).
Computational studies of the anharmonic vibrational properties of medium-size biomolecules have been limited mostly to VSCF\cite{Gregurick1997_VSCFPeptideWater,Gerber2002_CCVSCF} and VPT2\cite{Fornaro2015_UracilDimers,Fornaro2015_UracilSolidState} approaches, and only recently vibrational CI (VCI) studies were published mainly based on local mode approaches to reduce the computational effort.\cite{Jacob2009_LocalModes,Panek2014_LocalModes,Panek2016_AnharmonicBiomolecules} 
In this section, the convergence of the vibrational energies of {SarGly$^+$} is studied for a varying number of renormalized block states $m$.
In fact, if the value of $m$ required to reach convergence turns out to be small and largely independent of the system size, vDMRG can be a valuable alternative to local mode approaches for large systems. In this section we assess the performance of vDMRG for large systems and do not aim to accurately reproduce experimental frequencies. In fact, a quartic force field is usually insufficient for variational calculations and terms up to sixth order\cite{Crittenden2016_PyVCI} have to be included to obtain results that are close to their experimental counterparts. For vDMRG, a higher-order force-field simply requires longer strings of creation and annihilation operators and therefore increasing the scaling. However, such terms of the Hamiltonian can easily be provided in the input to our \textsc{QCMaquis-V} program. Furthermore, a more accurate electronic-structure method than the one employed here (B3LYP/6-311+G(d,p)) is necessary to reproduce experimental frequencies.

Due to the size of {SarGly$^+$}, theoretical results with fully-converged VCI calculations are not available.
However, a theoretical study of the vibrational properties of {SarGly$^+$} was performed recently with VSCF and a local mode ansatz based on a B3LYP/6-311+G(d,p) quartic force-field.\cite{Cheng2014_LocalModesVCI}
For the sake of comparability, the same electronic-structure approach was applied here for the structure optimization and the computation of the semi-quartic force field, where the quartic force constants $k_{ijkl}$ with four different indices were neglected together with those smaller than 1~cm$^{-1}$.
In view of the results of the previous sections, $\NMax$ was set to 6 and values of $m$ range from 5 to 20 in all calculations. 
The harmonic vibrational wave function served as the initial guess for the MPS. 
To limit the computational cost, the reduced-dimensionality (RD) scheme presented in Refs.~\citenum{Schuurman2005_ReducedDimensionality} and \citenum{Barone2013_Glycine} was exploited for treating all modes below 900 cm$^{-1}$ as harmonic ones, therefore neglecting all couplings between these modes and the fully anharmonically treated modes.
Such a selection of the modes might seem, in general, crude and prone to arbitrariness, and for larger systems a more accurate definition of the reduced-dimensionality model is certainly required.\cite{Kvapilova2015_RDMetals}
However, the aim of this section is to demonstrate the efficiency of vDMRG for large systems, not the accurate reproduction of experimental results that will also be limited by the electronic structure approach.
Therefore, an RD potential, consisting of 35 modes (corresponding to 35 DMRG sites), represents a viable setup for vDMRG.

A comparison of the RD and a full-dimensional scheme is presented in Table~S6 for generalized vibrational second-order perturbation theory (GVPT2, further details in Refs.~\citenum{Bloino2012_GVPT2} and \citenum{Bloino2016_Review}) calculations on {SarGly$^+$} and revealed that the energies deviate by less than 5~cm$^{-1}$ for most modes under 2000~cm$^{-1}$.
This indicates that the RD scheme is valid for this molecule within the desired accuracy. 
Obviously, the RD scheme largely improves the efficiency of the vDMRG calculation. 
In fact, the full-dimensional Hamiltonian would have a large number of low-frequency eigenvalues, corresponding to fundamentals, overtones, and combination bands of low-frequency modes, below the fingerprint region ($\approx 900-1700$~cm$^{-1}$). 

\def\arraystretch{1.25}
\begin{table*}[htb!]
  \begin{center}
  \caption{vDMRG vibrational transition energies calculated with Eq.~(\ref{trans_energy}) and zero-point energies (in cm$^{-1}$) for {SarGly$^+$} with $N_\text{max}=6$ and varying 
	       number of renormalized states $m$. Theoretical data ('l-VSCF', from Ref.~\citenum{Cheng2014_LocalModesVCI})
	       and experimental data ('Exp.', from Ref.~\citenum{Johnson2014_SarGlyExp}) are also reported. VPT2 data was calculated with \textsc{Gaussian}\cite{Bloino2012_GVPT2} for comparison.
  \label{tab:Energies_Dipeptide}                      
}
 \begin{tabular}{c|c|ccc|ccc}
    \hline\hline
       &  Harm     &   m=5    &   m=10   &   m=20   &    VPT2    & l-VSCF & Exp. \\
    \hline
  ZPVE & 39054.82  & 38601.40 & 38590.02 & 38572.25 &  38611.75  & &  \\
    1  &   987.26  &   961.71 &   961.14 &   958.04 &    973.56  & &  \\
    2  &  1011.54  &   988.95 &   985.83 &   986.50 &    992.70  & &  \\
    3  &  1038.64  &  1027.61 &  1029.15 &  1028.58 &   1038.57  & &  \\
    4  &  1100.19  &  1073.21 &  1075.45 &  1074.85 &   1084.04  & 1110 & 1088 \\
    5  &  1148.05  &  1116.81 &  1114.35 &  1112.45 &   1131.87  & 1169 & 1147 \\
    6  &  1187.57  &  1142.41 &  1141.26 &  1138.53 &   1162.39  & &  \\
    7  &  1190.81  &  1142.93 &  1142.99 &  1140.50 &   1174.23  & &  \\
    8  &  1242.55  &  1210.48 &  1210.82 &  1209.20 &   1216.47  & &  \\
    9  &  1257.74  &  1214.44 &  1214.36 &  1214.83 &   1232.04  & &  \\
    10 &  1287.33  &  1267.32 &  1245.69 &  1242.61 &   1263.71  & &  \\
    11 &  1317.02  &  1281.42 &  1276.75 &  1273.80 &   1287.62  & &  \\
    12 &  1337.91  &  1295.29 &  1295.79 &  1294.28 &   1309.18  & & \\
    13 &  1405.94  &  1344.23 &  1357.13 &  1350.49 &   1369.52  & &  \\
    14 &  1432.43  &  1358.03 &  1378.88 &  1375.89 &   1389.17  & 1394 & 1384 \\
    15 &  1435.01  &  1407.78 &  1404.80 &  1407.04 &   1418.96  & &  \\
    16 &  1464.25  &  1426.97 &  1420.19 &  1416.34 &   1432.58  & &  \\
    17 &  1484.39  &  1436.63 &  1431.33 &  1437.23 &   1445.66  & &  \\
    18 &  1491.29  &  1443.39 &  1443.08 &  1443.73 &   1453.07  & &  \\
    19 &  1500.94  &  1457.25 &  1455.59 &  1454.21 &   1458.22  & &  \\
    20 &  1505.58  &  1474.20 &  1467.00 &  1462.56 &   1466.01  & &  \\
    21 &  1565.36  &  1566.85 &  1566.85 &  1566.06 &   1538.98  & &  \\
    \hline\hline
  \end{tabular}                                                          
  \end{center}                       
\end{table*}

The vDMRG fundamental frequencies of {SarGly$^+$} in the fingerprint region are collected in Table~\ref{tab:Energies_Dipeptide}.
Localized VSCF (l-VSCF) from Ref.~\citenum{Cheng2014_LocalModesVCI} are given together with GVPT2 results obtained with the same RD potential as the one in the vDMRG calculations. 
Experimental frequencies from Ref.~\citenum{Johnson2014_SarGlyExp} are included where available. 
The convergence of the energies with respect to the number of renormalized block states $m$ is similar to that observed for smaller systems as analyzed in the previous sections. 
Almost fully converged results were obtained already with $m=10$ and corrections below 2~cm$^{-1}$ were observed upon increasing $m$ to 20. The ZPVE converges more slowly with respect to $m$ than all the excitation frequencies, with a difference of 8 cm$^{-1}$ between results obtained from $m$=10 and $m$=20. As shown in Figure S1 of the Supporting Information, converged results can be obtained with $m$=40, and a further increase of $m$ do not lead to any significant modifications. Due to error compensation for the transition energies, the convergence of the ZPVE appears to be slower. In the harmonic approximation, the contribution of high-energy modes to state energies is the same and equal to half of the fundamental frequency, and therefore this contribution cancels out when calculating excitation energies as states energy differences.

In Figure~\ref{fig:IR_dipeptide}, the spectrum obtained from vDMRG anharmonic frequencies of {SarGly$^+$} with $\NMax=6$ and $m=20$ and harmonic intensities is compared to the experimental spectrum.\cite{Johnson2014_SarGlyExp} 
As expected, the inclusion of anharmonic effects leads to an overall red shift of the frequencies, providing a considerably better agreement with the experimental data. 
This is pronounced for the band at 1147 cm$^{-1}$ ({C-O-H} bending), whose energy is overestimated by approximately 50~cm$^{-1}$ with harmonic calculations, whereas it is correctly reproduced by vDMRG. 
Similarly, the pattern recorded between 1350 and 1450 cm$^{-1}$, that is composed by three nearly equidistant bands with comparable intensity, is reproduced more accurately by vDMRG, whereas two of the three bands have nearly the same energy in the harmonic approximation. 
Therefore, although a potential higher than fourth order is usually required to obtain reliable variational energies, anharmonic variational calculations from a quartic potential considerably improve the description of this system compared to purely harmonic calculations.

These vDMRG calculations on {SarGly$^+$} with the largest set of parameters among the ones studied here ($\NMax=6$, $m=20$, 10~sweeps)  took, on average, 17.500 seconds per state (approximately 4 hours and 48 minutes). 
This corresponds to an overall computational time of less than four days for the lowest 20 excited states on an Intel Xeon E5-2670 @2.6 GHz with 2x8 central processing units and a 64 GB RAM node. 
Therefore, systems with more than 30 modes can be studied with manageable computational effort.

\begin{figure}[htb!]
	\centering
	\includegraphics[width=.49\textwidth]{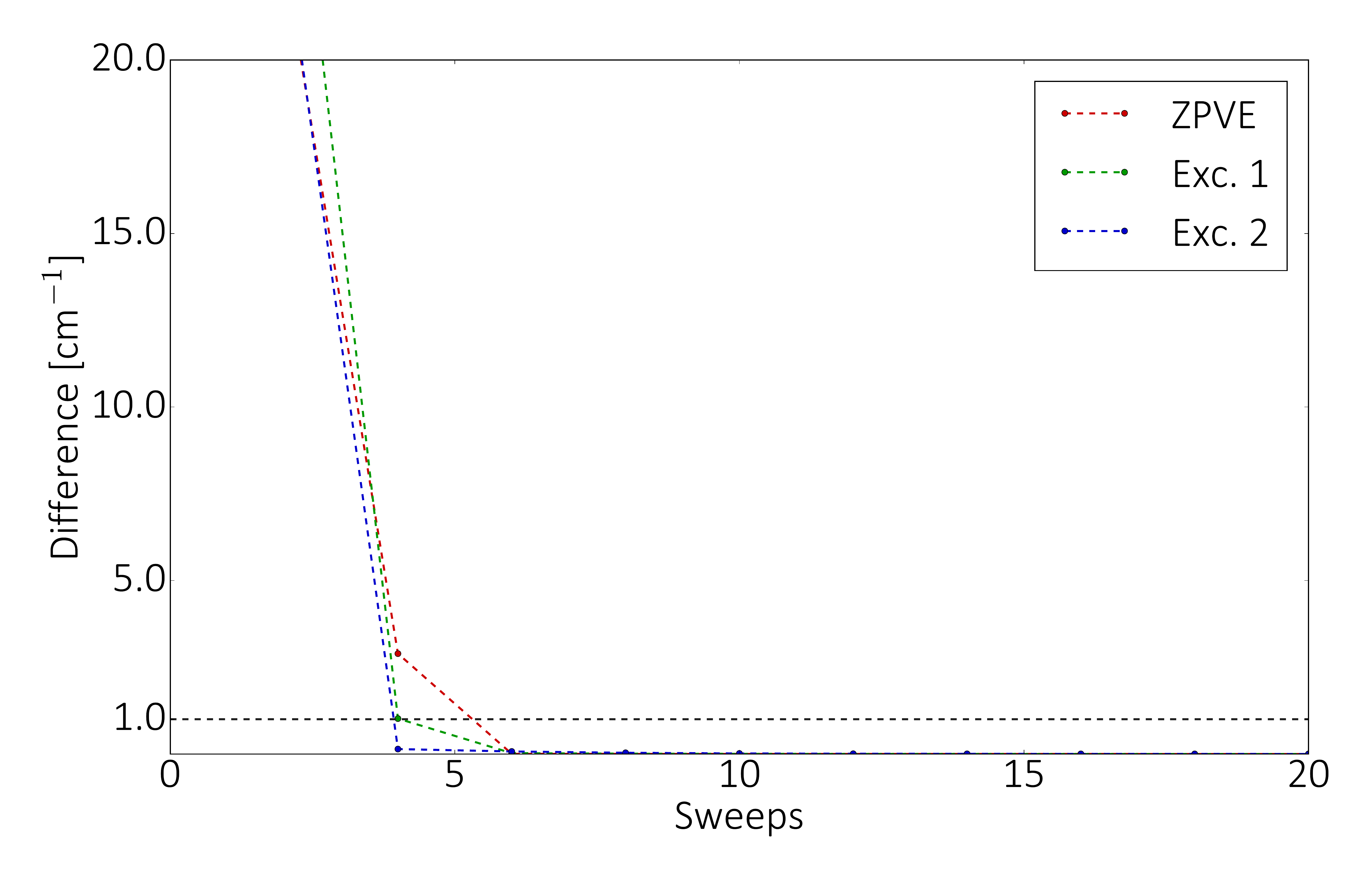}\\
	\includegraphics[width=.49\textwidth]{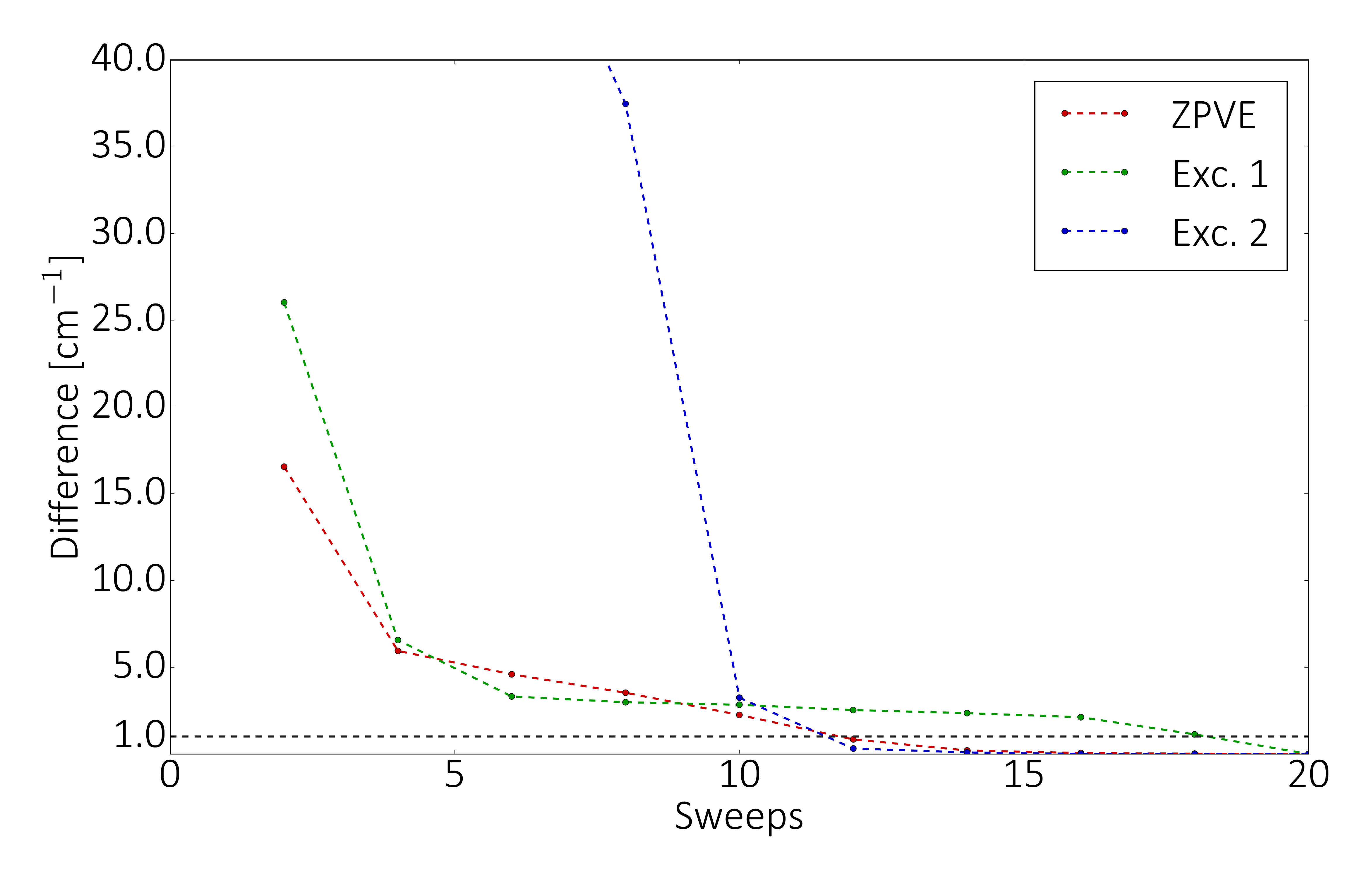}
	\caption{Energy of the ground and first two excited states of {C$_2$H$_4$} (upper panel) and {SarGly$^+$} (lower panel) as 
		     a function of the number of sweeps. In all cases, vDMRG calculations were performed with $\NMax=6$ and $m=10$. For {C$_2$H$_4$}, the full quartic potential from Ref.~\citenum{Crittenden2015_PyPES} was used. For {SarGly$^+$}, a semi-diagonal quartic  B3LYP/6-311+G(d,p) force-field was employed.
	\label{fig:Convergence}
}
\end{figure}

\begin{figure}[htb!]
	\centering
	\includegraphics[width=.49\textwidth]{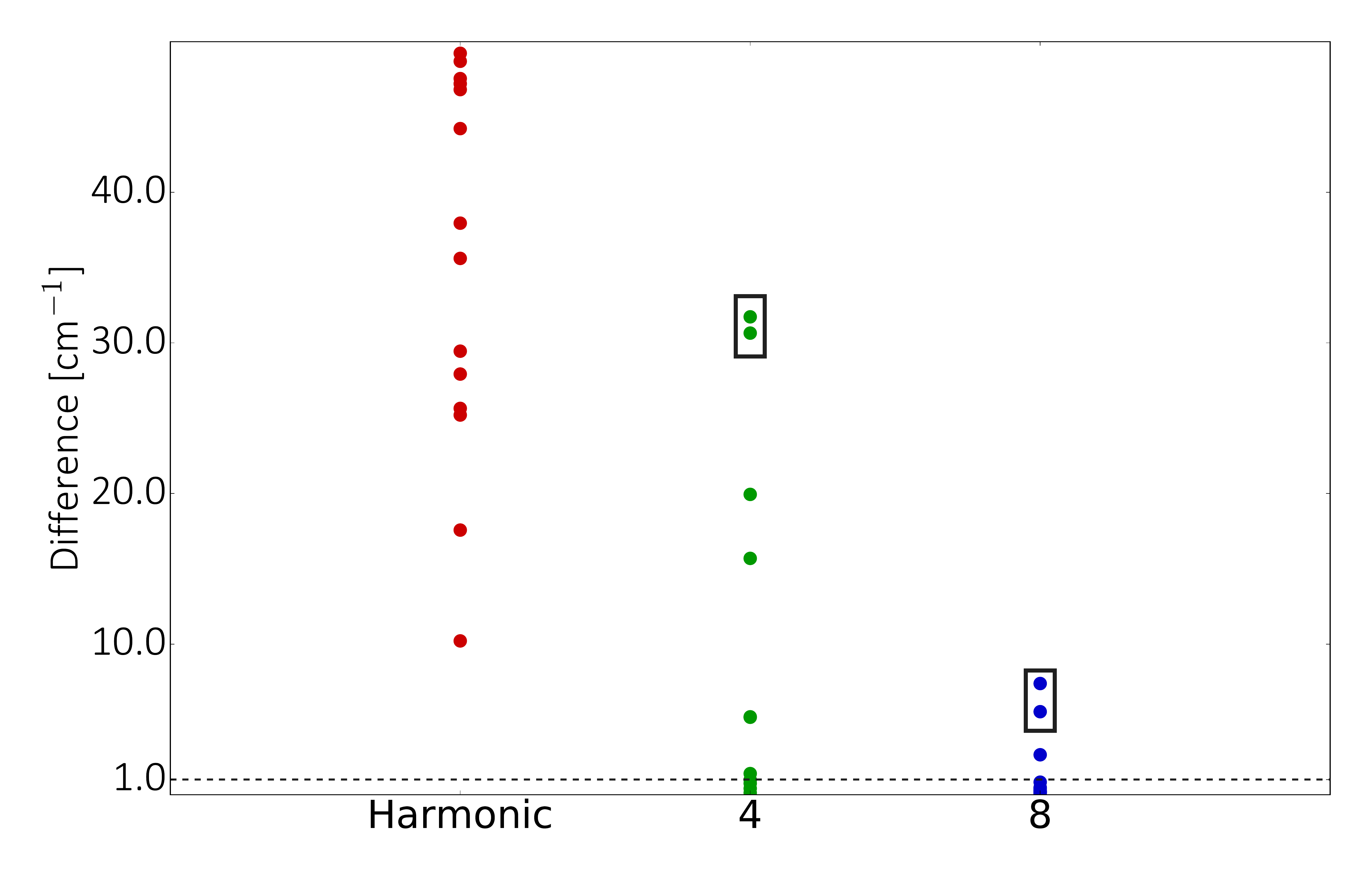}
	\caption{Difference between vibrational energies of {SarGly$^+$}, calculated with zero (Guess), 4 and 8 sweeps in the optimization procedure, 
	         and converged energies (obtained with 20 sweeps). vDMRG energies were calculated with a semi-diagonal quartic B3LYP/6-311+G(d,p) force-field. Data corresponding to modes 36 and 37 are highlighted with a black box.
	\label{fig:Convergence_overall}
}
\end{figure}

As already noted above, an additional factor determining the efficiency of the vDMRG calculation is the iterative optimization algorithm.
In fact, the number of renormalized block states $m$ and the dimension of the local basis $\NMax$ determine the computational cost of each sweep. 
However, in order to make vDMRG applicable to large molecules, the number of sweeps required to reach convergence should be largely independent of the system size.

In Figure~\ref{fig:Convergence}, the energies of the first three vibrational states of {SarGly$^+$} (lower panel) and {C$_2$H$_4$} (upper panel) are reported as a function of the number of sweeps.
The vDMRG calculations were performed with quartic force-fields (semi-diagonal for {SarGly$^+$}) with $\NMax=6$ and $m=10$.  
For ethylene, the convergence of the optimization algorithm is fast, with deviations below 1 cm$^{-1}$ from the converged value (obtained with 20 sweeps) already within six sweeps for both the ground and the excited states. 
The efficiency of the procedure is only slightly lower for {SarGly$^+$}, even if, to obtain a ground-state energy with an accuracy within 1~cm$^{-1}$, at least eight sweeps are required (eight for the first excited state). 

In Table S5 of the Supporting Information, the previous analysis has been extended to all vibrational energies below 1600 cm$^{-1}$. 
A graphical representation of the difference with respect to fully converged results as a function of the number of sweeps is reported in Figure~\ref{fig:Convergence_overall}. 
With only four sweeps, most of the vibrational frequencies are converged within 1~cm$^{-1}$, with the exception of only five frequencies.
With eight sweeps, only three frequencies were not converged which demonstrates the efficiency of the optimization procedure also for highly-excited states.
It is interesting to note that the frequencies that need the largest number of
sweeps to converge, $\nu_{36}$ and $\nu_{37}$, are involved in a Darling-Dennison 1-1 resonance (as can be tested by performing the test described, for example, in Ref.~\citenum{Bloino2015_ROAImplementation}) and, as noted above, anharmonic corrections are usually relevant for resonant states.
This indicates once more that the number of sweeps required to reach convergence increases with the magnitude of anharmonic corrections.
In order to increase the efficiency of the optimization algorithm in such cases, the GVPT2 vibrational states could be exploited as an initial guess of the optimization procedure, but this analysis is beyond the scope of the present work.

\section{Conclusions}                                                                    %

In this work, we presented the first theoretical formulation and implementation of an MPO-based vibrational DMRG algorithm for the calculation of vibrational properties of molecular systems.
The second-quantized Watson Hamiltonian was employed with a Taylor expansion up to sixth-order of the PES and Cartesian normal coordinates as the reference coordinate system.
This Hamiltonian was compactly represented as an MPO.

We demonstrated that highly-accurate, converged results can be obtained with vDMRG and a moderate number of renormalized block states.
However, the main advantage of vDMRG lies in the fact that results with an accuracy of 1-2 cm$^{-1}$ can be obtained with a very low number of renormalized block states also for large systems, with 30-40 normal modes.
Therefore, vDMRG represents a reliable method for the calculation of vibrational energies for large systems. In fact, very large variational space can be sampled without any restriction to a predefined excitation hierarchy. Furthermore, vDMRG could benefit from most of the strategies used for large-scale VCI calculations, such as state-selection algorithms and localized modes Hamiltonians.

Future work will focus on increasing the efficiency of vDMRG. 
First of all, the present implementation supports only the harmonic-oscillator wave functions as a local basis set for the individual modes.
However, it has already been shown for established approaches that other choices of the local basis set may provide a faster convergence of the variational expansion.
For example, basis functions arising from VSCF calculations are usually better suited for VCI calculations,\cite{Roy2013_VSCFReview} and the extension of our vDMRG implementation to support a VSCF reference is possible with the second-quantized Hamiltonian presented in Ref. \citenum{Christiansen2004_SecondQuantization}.
Being a one-dimensional algorithm, DMRG is most efficient for Hamiltonians with mainly short-range (or local) interactions. 
For this reason, the extension to localized modes,\cite{Jacob2009_LocalModes} possibly curvilinear,\cite{Arnim1999_GNIC,Changala2016_VMP2CurvCoord} is particularly appealing, especially for large systems.\cite{Jacob2009_PolypeptideLocalModes}
Furthermore, it might be beneficial to place resonant modes close to each other on the DMRG lattice and ordering methods based on quantum entanglement can be of value here.
An adaptation of the basis set size to the entanglement of a given mode with all other modes can additionally help to select a local basis with favorable convergence properties.

In the present work, vibrational frequencies were assigned to specific molecular vibrations by comparing the energies to reference data or to VPT2 results.
However, this procedure lacks robustness, especially for large systems with pronounced variational corrections. 
In order to build a more robust assignment algorithm, a sampling procedure already applied in electronic structure calculations to express an MPS in terms of a FCI basis set could be employed.\cite{Moritz2007_SlaterDecomposition,bogus11}

Finally, the full Hamiltonian is included in the variational treatment so far.
However, it has been shown\cite{Carbonniere2010_VCIP,Carbonniere2012_VCIP} that more efficient approaches can be devised, where only some terms of the Hamiltonian are included in the variational treatment, and the remaining terms are treated perturbatively.
The coupling of the vDMRG approach presented here with a perturbative treatment is possible in analogy to approaches already proposed for electronic structure methods\cite{shar14,ren16} and would reduce the size of the Hamiltonian that is treated variationally, and consequently also the computational cost.

\section*{Acknowledgments}
This work was supported by the Schweizerischer Nationalfonds (No. 20020\_169120). 
C.J.S. gratefully acknowledges a K\'ekule fellowship from the Fonds der Chemischen Industrie.
A.B. and V.B. acknowledge funding from the European Research Council under the European Unions Seventh Framework Programme (FP/2007-2013)/ERC Grant Agreement n.\ [320951] and the Italian MIUR (PRIN~2015 Grant Number 2015F59J3).

\section*{Supporting Information Available}
Additional information and results are collected in the supporting information.
This information is available free of charge via the Internet at http://pubs.acs.org/.

\section*{References}


\begin{thebibliography}{100}

\bibitem{IRBook}
Rijs,~A.~M.;\ \ Oomens,~J. \textit{{Gas-Phase IR Spectroscopy and Structure of
  Biological Molecules};} Springer International Publishing: 2015.

\bibitem{VCDBook}
Stephens,~P.~J.;\ \ Devlin,~F.;\ \ Cheeseman,~J.~R. \textit{{VCD spectroscopy
  for Organic Chemists};} CRC Press: 2012.

\bibitem{Nafie2014_RamanReview}
Nafie,~L.~A.  {Recent advances in linear and nonlinear Raman spectroscopy. Part
  VIII},  \textit{J. Raman Spectrosc.} \textbf{2014,} \textsl{45,} 1326-1346.

\bibitem{Bour2014_ROAReview}
Parchansky,~V.;\ \ Kapitan,~J.;\ \ Bour,~P.  {Inspecting chiral molecules by
  Raman optical activity spectroscopy},  \textit{RSC Adv.} \textbf{2014,}
  \textsl{4,} 57125-57136.

\bibitem{Barth2007_InfraredProtein}
Barth,~A.  {Infrared spectroscopy of proteins},  \textit{Biochim. Biophys.
  Acta} \textbf{2007,} \textsl{1767,} 1073 - 1101.

\bibitem{Herbst2009_ReviewAstro}
Herbst,~E.;\ \ van Dishoeck,~E.~F.  {Complex Organic Interstellar Molecules},
  \textit{Ann. Rev. Astron. Astrophys.} \textbf{2009,} \textsl{47,} 427-480.

\bibitem{Tielens2013_MolecularUniverse}
Tielens,~A. G. G.~M.  {The molecular universe},  \textit{Rev. Mod. Phys.}
  \textbf{2013,} \textsl{85,} 1021-1081.

\bibitem{Puzzarini2015_AccountAstro}
Barone,~V.;\ \ Biczysko,~M.;\ \ Puzzarini,~C.  {Quantum Chemistry Meets
  Spectroscopy for Astrochemistry: Increasing Complexity toward Prebiotic
  Molecules},  \textit{Acc. Chem. Res.} \textbf{2015,} \textsl{48,} 1413-1422.

\bibitem{Jilie2007_ReviewFTIR}
Kong,~J.;\ \ Yu,~S.  {Fourier Transform Infrared Spectroscopic Analysis of
  Protein Secondary Structures},  \textit{Acta Biochim. Biophys. Sin.}
  \textbf{2007,} \textsl{39,} 549-559.

\bibitem{Reiher2007_BioBook}
Herrmann,~C.;\ \ Reiher,~M.  {First-Principles Approach to Vibrational
  Spectroscopy of Biomolecules},  \textit{Top. Curr. Chem.} \textbf{2007,}
  \textsl{268,} 85-132.

\bibitem{Whitehead1975_VCIFirst}
Whitehead,~R.~J.;\ \ Handy,~N.~C.  {Variational calculation of
  vibration-rotation energy levels for triatomic molecules},  \textit{J. Mol.
  Spec.} \textbf{1975,} \textsl{55,} 356 - 373.

\bibitem{Romanowski1985_VCIFormaldehyde}
Romanowski,~H.;\ \ Bowman,~J.~M.;\ \ Harding,~L.~B.  {Vibrational energy levels
  of formaldehyde},  \textit{J. Chem. Phys.} \textbf{1985,} \textsl{82,}
  4155-4165.

\bibitem{Carter1986_VariationalVibrations}
Carter,~S.;\ \ Handy,~N.~C.  {The variational method for the calculation of
  Ro-vibrational energy levels},  \textit{Comput. Phys. Rep.} \textbf{1986,}
  \textsl{5,} 117 - 171.

\bibitem{Carbonniere2004_VCICoriolis}
Carbonni{\`e}re,~P.;\ \ Barone,~V.  {Coriolis couplings in variational
  computations of vibrational spectra beyond the harmonic approximation:
  implementation and validation},  \textit{Chem. Phys. Lett.} \textbf{2004,}
  \textsl{392,} 365-371.

\bibitem{Crittenden2016_PyVCI}
Sibaev,~M.;\ \ Crittenden,~D.~L.  {PyVCI: A flexible open-source code for
  calculating accurate molecular infrared spectra},  \textit{Comput. Phys.
  Commun.} \textbf{2016,} \textsl{203,} 290 - 297.

\bibitem{Carter1997_VCI}
Carter,~S.;\ \ Culik,~S.~J.;\ \ Bowman,~J.~M.  {Vibrational self-consistent
  field method for many-mode systems: A new approach and application to the
  vibrations of CO adsorbed on Cu(100)},  \textit{J. Chem. Phys.}
  \textbf{1997,} \textsl{107,} 10458-10469.

\bibitem{Gerber1999_VSCF}
Chaban,~G.~M.;\ \ Jung,~J.~O.;\ \ Gerber,~R.~B.  {Ab initio calculation of
  anharmonic vibrational states of polyatomic systems: Electronic structure
  combined with vibrational self-consistent field},  \textit{J. Chem. Phys.}
  \textbf{1999,} \textsl{111,} 1823-1829.

\bibitem{Gerber2002_CCVSCF}
Gerber,~R.~B.;\ \ Brauer,~B.;\ \ K.~Gregurick,~S.;\ \ M.~Chaban,~G.
  {Calculation of anharmonic vibrational spectroscopy of small biological
  molecules},  \textit{PhysChemComm} \textbf{2002,} \textsl{5,} 142-150.

\bibitem{Hirata2007_CO2}
Rodriguez-Garcia,~V.;\ \ Hirata,~S.;\ \ Yagi,~K.;\ \ Hirao,~K.;\ \
  Taketsugu,~T.;\ \ Schweigert,~I.;\ \ Tasumi,~M.  {Fermi resonance in CO$_2$:
  A combined electronic coupled-cluster and vibrational
  configuration-interaction prediction},  \textit{J. Chem. Phys.}
  \textbf{2007,} \textsl{126,} 124303.

\bibitem{Neff2009_LargeScaleVCI}
Neff,~M.;\ \ Rauhut,~G.  {Toward large scale vibrational configuration
  interaction calculations},  \textit{J. Chem. Phys.} \textbf{2009,}
  \textsl{131,} 124129.

\bibitem{chris07}
Christiansen,~O.  {Vibrational structure theory: new vibrational wave function
  methods for calculation of anharmonic vibrational energies and vibrational
  contributions to molecular properties},  \textit{Phys. Chem. Chem. Phys.}
  \textbf{2007,} \textsl{9,} 2942-2953.

\bibitem{scri08}
Scribano,~Y.;\ \ Benoit,~D.~M.  {Iterative active-space selection for
  vibrational configuration interaction calculations using a reduced-coupling
  VSCF basis},  \textit{Chem. Phys. Lett.} \textbf{2008,} \textsl{458,} 384
  - 387.

\bibitem{stro11}
Strobusch,~D.;\ \ Scheurer,~C.  The hierarchical expansion of the kinetic
  energy operator in curvilinear coordinates extended to the vibrational
  configuration interaction method,  \textit{J. Chem. Phys.} \textbf{2011,}
  \textsl{135,} 144101.

\bibitem{chris04}
Christiansen,~O.  {Vibrational coupled cluster theory},  \textit{J. Chem.
  Phys.} \textbf{2004,} \textsl{120,} 2149-2159.

\bibitem{Mills1961_VPT2First}
Mills,~I.  {Vibrational perturbation theory},  \textit{J. Mol. Spec.}
  \textbf{1961,} \textsl{5,} 334 - 340.

\bibitem{gerb05}
Gerber,~R.;\ \ Chaban,~G.;\ \ Brauer,~B.;\ \ Miller,~Y.  First-principles
  calculations of anharmonic vibrational spectroscopy of large molecules.   In
  \textit{Theory and Applications of Computational Chemistry}; Dykstra,~C.~E.;\
  \ Frenking,~G.;\ \ Kim,~K.~S.;\ \ Scuseria,~G.~E.,\ \ Eds.;  Elsevier:
  Amsterdam, 2005.

\bibitem{Barone2005_VPT2}
Barone,~V.  {Anharmonic vibrational properties by a fully automated
  second-order perturbative approach},  \textit{J. Chem. Phys.} \textbf{2005,}
  \textsl{122,} 014108.

\bibitem{Krasnoshchekov2014_VVPT2}
Krasnoshchekov,~S.~V.;\ \ Isayeva,~E.~V.;\ \ Stepanov,~N.~F.  {Criteria for
  first- and second-order vibrational resonances and correct evaluation of the
  Darling-Dennison resonance coefficients using the canonical Van Vleck
  perturbation theory},  \textit{J. Chem. Phys.} \textbf{2014,} \textsl{141,}
  234114.

\bibitem{Rosnik2014_VPT2K}
Rosnik,~A.~M.;\ \ Polik,~W.~F.  {VPT2+K spectroscopic constants and matrix
  elements of the transformed vibrational Hamiltonian of a polyatomic molecule
  with resonances using Van Vleck perturbation theory},  \textit{Mol. Phys.}
  \textbf{2014,} \textsl{112,} 261-300.

\bibitem{Norris1996_VMP2First}
Norris,~L.~S.;\ \ Ratner,~M.~A.;\ \ Roitberg,~A.~E.;\ \ Gerber,~R.~B.
  {M\"{o}ller-Plesset perturbation theory applied to vibrational problems},
  \textit{J. Chem. Phys.} \textbf{1996,} \textsl{105,} 11261-11267.

\bibitem{Christiansen2003_VMP2}
Christiansen,~O.  {M{\o}ller-Plesset perturbation theory for vibrational wave
  functions},  \textit{J. Chem. Phys.} \textbf{2003,} \textsl{119,} 5773-5781.

\bibitem{Changala2016_VMP2CurvCoord}
Changala,~P.~B.;\ \ Baraban,~J.~H.  {Ab initio effective rotational and
  rovibrational Hamiltonians for non-rigid systems via curvilinear second order
  vibrational M{\o}ller--Plesset perturbation theory},  \textit{J. Chem. Phys.}
  \textbf{2016,} \textsl{145,} 174106.

\bibitem{Dawes2005_BasisPruning}
Dawes,~R.;\ \ Carrington~Jr.,~T.  {How to choose one-dimensional basis
  functions so that a very efficient multidimensional basis may be extracted
  from a direct product of the one-dimensional functions: Energy levels of
  coupled systems with as many as 16 coordinates},  \textit{J. Chem. Phys.}
  \textbf{2005,} \textsl{122,} 134101.

\bibitem{Avila2011_C2H4PrunedBasis}
Avila,~G.;\ \ Carrington~Jr.,~T.  {Using a pruned basis, a non-product
  quadrature grid, and the exact Watson normal-coordinate kinetic energy
  operator to solve the vibrational Schr\"{o}dinger equation for C$_2$H$_4$},
  \textit{J. Chem. Phys.} \textbf{2011,} \textsl{135,} 064101.

\bibitem{Bramley1993_Contraction}
Bramley,~M.~J.;\ \ Handy,~N.~C.  {Efficient calculation of rovibrational
  eigenstates of sequentially bonded four--atom molecules},  \textit{J. Chem.
  Phys.} \textbf{1993,} \textsl{98,} 1378-1397.

\bibitem{Wang2002_ContractedBasis}
Wang,~X.-G.;\ \ Carrington~Jr.,~T.  {New ideas for using contracted basis
  functions with a Lanczos eigensolver for computing vibrational spectra of
  molecules with four or more atoms},  \textit{J. Chem. Phys.} \textbf{2002,}
  \textsl{117,} 6923-6934.

\bibitem{Jacob2009_LocalModes}
Jacob,~C.~R.;\ \ Reiher,~M.  {Localizing normal modes in large molecules},
  \textit{J. Chem. Phys.} \textbf{2009,} \textsl{130,} 084106.

\bibitem{Jacob2009_PolypeptideLocalModes}
Jacob,~C.~R.;\ \ Luber,~S.;\ \ Reiher,~M.  {Analysis of Secondary Structure
  Effects on the IR and Raman Spectra of Polypeptides in Terms of Localized
  Vibrations},  \textit{J. Phys. Chem. B} \textbf{2009,} \textsl{113,}
  6558-6573.

\bibitem{Panek2014_LocalModes}
Panek,~P.~T.;\ \ Jacob,~C.~R.  {Efficient Calculation of Anharmonic Vibrational
  Spectra of Large Molecules with Localized Modes},  \textit{ChemPhysChem}
  \textbf{2014,} \textsl{15,} 3365-3377.

\bibitem{Cheng2014_LocalModesVCI}
Cheng,~X.;\ \ Steele,~R.~P.  {Efficient anharmonic vibrational spectroscopy for
  large molecules using local-mode coordinates},  \textit{J. Chem. Phys.}
  \textbf{2014,} \textsl{141,} 104105.

\bibitem{klin15}
Klinting,~E.~L.;\ \ K\"{o}nig,~C.;\ \ Christiansen,~O.  {Hybrid Optimized and
  Localized Vibrational Coordinates},  \textit{J. Phys. Chem. A} \textbf{2015,}
  \textsl{119,} 11007-11021.

\bibitem{pane16}
Panek,~P.~T.;\ \ Jacob,~C.~R.  On the benefits of localized modes in anharmonic
  vibrational calculations for small molecules,  \textit{J. Chem. Phys.}
  \textbf{2016,} \textsl{144,} 164111.

\bibitem{lecl16a}
Leclerc,~A.;\ \ Thomas,~P.~S.;\ \ Carrington,~T.  {Comparison of different
  eigensolvers for calculating vibrational spectra using low-rank,
  sum-of-product basis functions},  \textit{Mol. Phys.} \textbf{2016,}
  accepted, doi:10.1080/00268976.2016.1249980.

\bibitem{whit92}
White,~S.~R.  {Density Matrix Formulation for Quantum Renormalization Groups},
  \textit{Phys. Rev. Lett.} \textbf{1992,} \textsl{69,} 2863--2866.

\bibitem{whit93}
White,~S.~R.  {Density-Matrix Algorithms for Quantum Renormalization Groups},
  \textit{Phys. Rev. B} \textbf{1993,} \textsl{48,} 10345--10356.

\bibitem{Schollwoeck2005}
Schollw\"ock,~U.  {The density-matrix renormalization group},  \textit{Rev.
  Mod. Phys.} \textbf{2005,} \textsl{77,} 259-315.

\bibitem{lege08}
Legeza,~{\"O}.;\ \ Noack,~R.;\ \ S\'olyom,~J.;\ \ Tincani,~L.  {Applications of
  Quantum Information in the Density-Matrix Renormalization Group},
  \textit{Lect. Notes Phys.} \textbf{2008,} \textsl{739,} 653-664.

\bibitem{chan08}
Chan,~G. K.-L.;\ \ Dorando,~J.~J.;\ \ Ghosh,~D.;\ \ Hachmann,~J.;\ \
  Neuscamman,~E.;\ \ Wang,~H.;\ \ Yanai,~T.  {An Introduction to the Density
  Matrix Renormalization Group Ansatz in Quantum Chemistry},  \textit{Prog.
  Theor. Chem. Phys.} \textbf{2008,} \textsl{18,} 49-65.

\bibitem{chan09}
Chan,~G. K.-L.;\ \ Zgid,~D.  {The Density Matrix Renormalization Group in
  Quantum Chemistry},  \textit{Annu. Rep. Comput. Chem.} \textbf{2009,}
  \textsl{5,} 149-162.

\bibitem{mart10}
Marti,~K.~H.;\ \ Reiher,~M.  {The Density Matrix Renormalization Group
  Algorithm in Quantum Chemistry},  \textit{Z. Phys. Chem.} \textbf{2010,}
  \textsl{224,} 583-599.

\bibitem{mart11}
Marti,~K.~H.;\ \ Reiher,~M.  {New Electron Correlation Theories for Transition
  Metal Chemistry},  \textit{Phys. Chem. Chem. Phys.} \textbf{2011,}
  \textsl{13,} 6750-6759.

\bibitem{chan11}
Chan,~G. K.-L.;\ \ Sharma,~S.  {The Density Matrix Renormalization Group in
  Chemistry},  \textit{Ann. Rev. Phys. Chem.} \textbf{2011,} \textsl{62,} 465.

\bibitem{Schollwoeck2011}
Schollw\"{o}ck,~U.  {The density-matrix renormalization group in the age of
  matrix product states},  \textit{Ann. Phys.} \textbf{2011,} \textsl{326,} 96
  - 192.

\bibitem{kura14}
Kurashige,~Y.  {Multireference Electron Correlation Methods with Density Matrix
  Renormalisation Group Reference Functions},  \textit{Mol. Phys.}
  \textbf{2014,} \textsl{112,} 1485-1494.

\bibitem{wout14}
Wouters,~S.;\ \ Van~Neck,~D.  {The Density Matrix Renormalization Group for ab
  initio Quantum Chemistry},  \textit{Eur. Phys. J. D} \textbf{2014,}
  \textsl{68,} 272.

\bibitem{yana15}
Yanai,~T.;\ \ Kurashige,~Y.;\ \ Mizukami,~W.;\ \ Chalupsk{\'y},~J.;\ \
  Lan,~T.~N.;\ \ Saitow,~M.  {Density Matrix Renormalization Group for ab
  initio Calculations and Associated Dynamic Correlation Methods: A Review of
  Theory and Applications},  \textit{Int. J. Quantum Chem.} \textbf{2015,}
  \textsl{115,} 283-299.

\bibitem{szal15}
Szalay,~S.;\ \ Pfeffer,~M.;\ \ Murg,~V.;\ \ Barcza,~G.;\ \ Verstraete,~F.;\ \
  Schneider,~R.;\ \ Legeza,~{\"O}.  {Tensor Product Methods and Entanglement
  Optimization for ab initio Quantum Chemistry},  \textit{Int. J. Quantum
  Chem.} \textbf{2015,} \textsl{115,} 1342-1391.

\bibitem{knec16}
Knecht,~S.;\ \ Hedeg{\aa}rd,~E.~D.;\ \ Keller,~S.;\ \ Kovyrshin,~A.;\ \
  Ma,~Y.;\ \ Muolo,~A.;\ \ Stein,~C.~J.;\ \ Reiher,~M.  {New Approaches for ab
  initio Calculations of Molecules with Strong Electron Correlation},
  \textit{Chimia} \textbf{2016,} \textsl{70,} 244-251.

\bibitem{chan16}
Chan,~G. K.-L.;\ \ Keselman,~A.;\ \ Nakatani,~N.;\ \ Li,~Z.;\ \ White,~S.~R.
  {Matrix Product Operators, Matrix Product States, and ab initio Density
  Matrix Renormalization Group Algorithms},  \textit{J. Chem. Phys.}
  \textbf{2016,} \textsl{145,} 014102.

\bibitem{Rakhuba2016_TensorTrain}
Rakhuba,~M.;\ \ Oseledets,~I.  {Calculating vibrational spectra of molecules
  using tensor train decomposition},  \textit{J. Chem. Phys.} \textbf{2016,}
  \textsl{145,} 124101.

\bibitem{Colbert1992_DVR}
Colbert,~D.~T.;\ \ Miller,~W.~H.  {A novel discrete variable representation for
  quantum mechanical reactive scattering via the S-matrix Kohn method},
  \textit{J. Chem. Phys.} \textbf{1992,} \textsl{96,} 1982-1991.

\bibitem{Carbonniere2005_CH3CN}
Begue,~D.;\ \ Carbonni\`{e}re,~P.;\ \ Pouchan,~C.  {Calculations of Vibrational
  Energy Levels by Using a Hybrid ab Initio and DFT Quartic Force Field:
  Application to Acetonitrile},  \textit{J. Phys. Chem. A} \textbf{2005,}
  \textsl{109,} 4611-4616.

\bibitem{Delahaye2014_EthylenePES}
Delahaye,~T.;\ \ Nikitin,~A.;\ \ Rey,~M.;\ \ Szalay,~P.;\ \ Tyuterev,~V.~G.  {A
  new accurate ground-state potential energy surface of ethylene and
  predictions for rotational and vibrational energy levels},  \textit{J. Chem.
  Phys.} \textbf{2014,} \textsl{141,} 104301.

\bibitem{Georges1999_C2H4Exp}
Georges,~R.;\ \ Bach,~M.;\ \ Herman,~M.  {The vibrational energy pattern in
  ethylene ($^{12}$C$_2$H$_4$)},  \textit{Mol. Phys.} \textbf{1999,}
  \textsl{97,} 279-292.

\bibitem{aqui8}
Aquilante,~F. \textit{et al.}\   {Molcas 8: New Capabilities for
  Multiconfigurational Quantum Chemical Calculations Across the Periodic
  Table},  \textit{J. Comput. Chem.} \textbf{2016,} \textsl{37,} 506-541.

\bibitem{wats68}
Watson,~J.~K.  {Simplification of the molecular vibration-rotation
  hamiltonian},  \textit{Mol. Phys.} \textbf{1968,} \textsl{15,} 479-490.

\bibitem{Hirata2014_SecondQuantization}
Hirata,~S.;\ \ Hermes,~M.~R.  {Normal-ordered second-quantized Hamiltonian for
  molecular vibrations},  \textit{J. Chem. Phys.} \textbf{2014,} \textsl{141,}.

\bibitem{Christiansen2004_SecondQuantization}
Christiansen,~O.  {A second quantization formulation of multimode dynamics},
  \textit{J. Chem. Phys.} \textbf{2004,} \textsl{120,} 2140-2148.

\bibitem{Wang2009_SQMCTDH}
Wang,~H.;\ \ Thoss,~M.  {Numerically exact quantum dynamics for
  indistinguishable particles: The multilayer multiconfiguration time-dependent
  Hartree theory in second quantization representation},  \textit{J. Chem.
  Phys.} \textbf{2009,} \textsl{131,} 024114.

\bibitem{Bowman2003_multimode}
Bowman,~J.M.;\ \ Carter,S.~;\ \ Huang,X.~  {MULTIMODE: A code to calculate 
	rovibrational energies of polyatomic molecules},  \textit{Int. Rev. Phys. Chem.} 
    \textbf{2003,} \textsl{22,} 533-549.

\bibitem{Gerber1995_vscf}
Roitberg,~A.;\ \ Gerber,R.B.~;\ \ Elber,R.~;\ \ Ratner,M.A.~ {Anharmonic wave functions 
	of proteins: quantum self-consistent field calculations of BPTI.},  \textit{Science} 
	\textbf{1995,} \textsl{268,} 1319-1322.

\bibitem{Keller2015}
Keller,~S.;\ \ Dolfi,~M.;\ \ Troyer,~M.;\ \ Reiher,~M.  {An efficient matrix
  product operator representation of the quantum chemical Hamiltonian},
  \textit{J. Chem. Phys.} \textbf{2015,} \textsl{143,}.

\bibitem{bau11}
Bauer,~B. \textit{et al.}\   The ALPS project release 2.0: open source software
  for strongly correlated systems,  \textit{J. Stat. Mech.} \textbf{2011,}
  \textsl{2011,} P05001.

\bibitem{Dolfi2014_ALPSImplementation}
Dolfi,~M.;\ \ Bauer,~B.;\ \ Keller,~S.;\ \ Kosenkov,~A.;\ \ Ewart,~T.;\ \
  Kantian,~A.;\ \ Giamarchi,~T.;\ \ Troyer,~M.  {Matrix product state
  applications for the ALPS project},  \textit{Comput. Phys. Commun.}
  \textbf{2014,} \textsl{185,} 3430 - 3440.

\bibitem{g16.a03}
Frisch,~M.~J. \textit{et al.}\  ``{Gaussian 16 {R}evision "{A}.03}'',  2016
  Gaussian Inc. Wallingford CT.

\bibitem{Peterson1998_Variational}
Peterson,~K.~A.  {Accurate ab initio near-equilibrium potential energy and
  dipole moment functions of the X $^2$B$_1$ and first excited $^2$A$_2$
  electronic states of OClO and OBrO},  \textit{J. Chem. Phys.} \textbf{1998,}
  \textsl{109,} 8864-8875.

\bibitem{Crittenden2015_PyPES}
Sibaev,~M.;\ \ Crittenden,~D.~L.  {The PyPES library of high quality
  semi-global potential energy surfaces},  \textit{J. Comput. Chem.}
  \textbf{2015,} \textsl{36,} 2200-2207.

\bibitem{Ortigoso1991_IRCl2OFirstBand}
Ortigoso,~J.;\ \ Escribano,~R.;\ \ Burkholder,~J.~B.;\ \ Howard,~C.~J.;\ \
  Lafferty,~W.~J.  {High-resolution infrared spectrum of the $\nu_1$ band of
  OClO},  \textit{J. Mol. Spectrosc.} \textbf{1991,} \textsl{148,} 346-370.

\bibitem{Ortigoso1992_IRCl2OOtherBands}
Ortigoso,~J.;\ \ Escribano,~R.;\ \ Burkholder,~J.~B.;\ \ Lafferty,~W.~J.  {The
  $\nu_2$ and $\nu_3$ bands and ground state constants of OClO},  \textit{J.
  Mol. Spectrosc.} \textbf{1992,} \textsl{155,} 25-43.

\bibitem{Ortigoso1993_IRCl2OCombBands}
Ortigoso,~J.;\ \ Escribano,~R.;\ \ Burkholder,~J.~B.;\ \ Lafferty,~W.~J.
  {Infrared Spectrum of OClO in the 2000 cm$^{-1}$ region: The 2$\nu_1$ and
  $\nu_1$ + $\nu_3$ Bands},  \textit{J. Mol. Spectrosc.} \textbf{1993,}
  \textsl{158,} 347-356.

\bibitem{Bacic1989_DVR}
Bacic,~Z.;\ \ Light,~J.~C.  {Theoretical Methods for Rovibrational States of
  Floppy Molecules},  \textit{Annu. Rev. Phys. Chem.} \textbf{1989,}
  \textsl{40,} 469-498.

\bibitem{Hoy1972_Tensor}
Hoy,~A.;\ \ Mills,~I.;\ \ Strey,~G.  {Anharmonic force constant calculations},
  \textit{Mol. Phys.} \textbf{1972,} \textsl{24,} 1265-1290.

\bibitem{begu05}
Begue,~D.;\ \ Carbonniere,~P.;\ \ Pouchan,~C.  {Calculations of Vibrational
  Energy Levels by Using a Hybrid ab Initio and DFT Quartic Force Field:
  Application to Acetonitrile},  \textit{J.Phys. Chem. A} \textbf{2005,}
  \textsl{109,} 4611-4616.

\bibitem{avil11}
Avila,~G.;\ \ Carrington~Jr.,~T.  {Using nonproduct quadrature grids to solve
  the vibrational Schr\"{o}dinger equation in 12D},  \textit{J. Chem. Phys.}
  \textbf{2011,} \textsl{134,} 054126.

\bibitem{lecl14}
Leclerc,~A.;\ \ Carrington,~T.  {Calculating vibrational spectra with sum of
  product basis functions without storing full-dimensional vectors or
  matrices},  \textit{J. Chem. Phys.} \textbf{2014,} \textsl{140,} 174111.

\bibitem{lecl16}
Leclerc,~A.;\ \ Carrington~Jr.,~T.  {Using symmetry-adapted optimized
  sum-of-products basis functions to calculate vibrational spectra},
  \textit{Chem. Phys. Lett.} \textbf{2016,} \textsl{644,} 183 - 188.

\bibitem{henr61}
Henry,~L.;\ \ Amat,~G.  {The cubic anharmonic potential function of polyatomic
  molecules},  \textit{J. Mol. Spectrosc.} \textbf{1961,} \textsl{5,} 319 -
  325.

\bibitem{henr65}
Henry,~L.;\ \ Amat,~G.  {The quartic anharmonic potential function of
  polyatomic molecules},  \textit{J. Mol. Spectrosc.} \textbf{1965,}
  \textsl{15,} 168 - 179.

\bibitem{picc15}
Piccardo,~M.;\ \ Bloino,~J.;\ \ Barone,~V.  {Generalized vibrational
  perturbation theory for rotovibrational energies of linear, symmetric and
  asymmetric tops: Theory, approximations, and automated approaches to deal
  with medium-to-large molecular systems},  \textit{Int. J. Quantum Chem.}
  \textbf{2015,} \textsl{115,}.

\bibitem{Moritz2007_SlaterDecomposition}
Moritz,~G.;\ \ Reiher,~M.  {Decomposition of density matrix renormalization
	group states into a Slater determinant basis},  \textit{J. Chem. Phys.}
\textbf{2007,} \textsl{126,} 244109.

\bibitem{bogus11}
Boguslawski,~K.;\ \ Marti,~K.~H.;\ \ Reiher,~M.  {Construction of CASCI-type
	wave functions for very large active spaces},  \textit{J. Chem. Phys.}
\textbf{2011,} \textsl{134,} 224101.

\bibitem{coriani12}
Coriani,~S.;\ \ Christiansen,~O.;\ \ Fransson,~T.;\ \ Norman,~P.  {Coupled-cluster 
	response theory for near-edge x-ray-absorption fine structure of atoms and molecules},  
	\textit{Phys. Rev. A} \textbf{2012,} \textsl{85,} 022507.
	
\bibitem{coriani13}
Fransson,~T.;\ \ Coriani,~S.;\ \ Christiansen,~O.;\ \ Norman,~P.  {Carbon X-ray 
	absorption spectra of fluoroethenes and acetone: A study at the coupled cluster, 
	density functional, and static-exchange levels of theory },  
\textit{J. Chem. Phys.} \textbf{2013,} \textsl{138,} 124311.

\bibitem{chan07}
Dorando,~J.J.;\ \ Hachmann,~J.;\ \ Chan,~G. K.-L. {Targeted excited state algorithm},  
\textit{J. Chem. Phys.} \textbf{2007,} \textsl{127,} 084109.

\bibitem{li11}
Liang,~W.;\ \ Fischer,~S.A.;\ \ Frisch,~M. J.;\ \ Li,~X. {Energy-Specific Linear Response 
	TDHF/TDDFT for Calculating High-Energy Excited States},  
\textit{J. Chem. Theo. Comput.} \textbf{2011,} \textsl{7,} 3540-3547.

\bibitem{petrenko17}
Petrenko,~T.;\ \ Rauhut,~G.; {A new efficient method for the calculation of interior 
	eigenpairs and its application to vibrational structure problems},  
\textit{J. Chem. Phys.} \textbf{2017,} \textsl{146,} 124101.

\bibitem{neugebauer11}
Kovyrshin,~A.;\ \ Neugebauer,~J.; {Potential-energy surfaces of local excited states 
	from subsystem- and selective Kohn–Sham-TDDFT},  
\textit{Chem. Phys.} \textbf{2011,} \textsl{391,} 147-156.

\bibitem{neugebauer03}
Reiher,~M.;\ \ Neugebauer,~J.; {A mode-selective quantum chemical method for tracking 
	molecular vibrations applied to functionalized carbon nanotubes},  
\textit{J. Chem. Phys.} \textbf{2003,} \textsl{118,} 1634.

\bibitem{Martin1995_C2H4Potential}
Martin,~J. M.~L.;\ \ Lee,~T.~J.;\ \ Taylor,~P.~R.;\ \ Francois,~J.-P.  {The
  anharmonic force field of ethylene, C$_2$H$_4$, by means of accurate ab
  initio calculations},  \textit{J. Chem. Phys.} \textbf{1995,} \textsl{103,}
  2589-2602.

\bibitem{Carter2012_C2H4Multimode}
Carter,~S.;\ \ Sharma,~A.~R.;\ \ Bowman,~J.~M.  {First-principles calculations
  of rovibrational energies, dipole transition intensities and partition
  function for ethylene using MULTIMODE},  \textit{J. Chem. Phys.}
  \textbf{2012,} \textsl{137,} 154301.

\bibitem{Bloino2012_GVPT2}
Bloino,~J.;\ \ Barone,~V.  {A second-order perturbation theory route to
  vibrational averages and transition properties of molecules: General
  formulation and application to infrared and vibrational circular dichroism
  spectroscopies},  \textit{J. Chem. Phys.} \textbf{2012,} \textsl{136,}
  124108.

\bibitem{Gregurick1997_VSCFPeptideWater}
Gregurick,~S.~K.;\ \ Fredj,~E.;\ \ Elber,~R.;\ \ Gerber,~R.~B.  {Vibrational
  Spectroscopy of Peptides and Peptide--Water Complexes: Anharmonic
  Coupled-Mode Calculations},  \textit{J. Phys. Chem. B} \textbf{1997,}
  \textsl{101,} 8595-8606.

\bibitem{Fornaro2015_UracilDimers}
Fornaro,~T.;\ \ Burini,~D.;\ \ Biczysko,~M.;\ \ Barone,~V.  {Hydrogen-Bonding
  Effects on Infrared Spectra from Anharmonic Computations: Uracil--Water
  Complexes and Uracil Dimers},  \textit{J. Phys. Chem. A} \textbf{2015,}
  \textsl{119,} 4224-4236.

\bibitem{Fornaro2015_UracilSolidState}
Fornaro,~T.;\ \ Carnimeo,~I.;\ \ Biczysko,~M.  {Toward Feasible and
  Comprehensive Computational Protocol for Simulation of the Spectroscopic
  Properties of Large Molecular Systems: The Anharmonic Infrared Spectrum of
  Uracil in the Solid State by the Reduced Dimensionality/Hybrid VPT2
  Approach},  \textit{J. Phys. Chem. A} \textbf{2015,} \textsl{119,} 5313-5326.

\bibitem{Panek2016_AnharmonicBiomolecules}
Panek,~P.~T.;\ \ Jacob,~C.~R.  {Anharmonic Theoretical Vibrational Spectroscopy
  of Polypeptides},  \textit{J. Phys. Chem. Lett.} \textbf{2016,} \textsl{7,}
  3084-3090.

\bibitem{Schuurman2005_ReducedDimensionality}
Schuurman,~M.~S.;\ \ Allen,~W.~D.;\ \ von Schleyer,~P.;\ \ {Schaefer
  III},~H.~F.  {The highly anharmonic BH$_5$ potential energy surface
  characterized in the ab initio limit},  \textit{J. Chem. Phys.}
  \textbf{2005,} \textsl{122,} 104302.

\bibitem{Barone2013_Glycine}
Barone,~V.;\ \ Biczysko,~M.;\ \ Bloino,~J.;\ \ Puzzarini,~C.  {Characterization
  of the Elusive Conformers of Glycine from State-of-the-Art Structural,
  Thermodynamic, and Spectroscopic Computations: Theory Complements
  Experiment},  \textit{J. Chem. Theory Comput.} \textbf{2013,} \textsl{9,}
  1533-1547.

\bibitem{Kvapilova2015_RDMetals}
Kvapilova,~H.;\ \ Vlcek,~A.;\ \ Barone,~V.;\ \ Biczysko,~M.;\ \ Zalis,~S.
  {Anharmonicity Effects in IR Spectra of [Re(X)(CO)$_3$($\alpha$--diimine)]
  ($\alpha$--diimine = 2,2--bipyridine or pyridylimidazo[1,5-a]pyridine; X = Cl
  or NCS) Complexes in Ground and Excited Electronic States},  \textit{J. Phys.
  Chem. A} \textbf{2015,} \textsl{119,} 10137--10146.

\bibitem{Bloino2016_Review}
Bloino,~J.;\ \ Baiardi,~A.;\ \ Biczysko,~M.  {Aiming at an accurate prediction
  of vibrational and electronic spectra for medium-to-large molecules: An
  overview},  \textit{Int. J. Quantum Chem.} \textbf{2016,} \textsl{116,}
  1543-1574.

\bibitem{Johnson2014_SarGlyExp}
Johnson,~C.~J.;\ \ Wolk,~A.~B.;\ \ Fournier,~J.~A.;\ \ Sullivan,~E.~N.;\ \
  Weddle,~G.~H.;\ \ Johnson,~M.~A.  {Communication: He-tagged vibrational
  spectra of the SarGlyH$^+$ and H$^+$(H$_2$O)$_{2,3}$ ions: Quantifying tag
  effects in cryogenic ion vibrational predissociation (CIVP) spectroscopy},
  \textit{J. Chem. Phys.} \textbf{2014,} \textsl{140,} 221101.

\bibitem{Bloino2015_ROAImplementation}
Bloino,~J.;\ \ Biczysko,~M.;\ \ Barone,~V.  {Anharmonic Effects on Vibrational
  Spectra Intensities: Infrared, Raman, Vibrational Circular Dichroism, and
  Raman Optical Activity},  \textit{J. Phys. Chem. A} \textbf{2015,}
  \textsl{119,} 11862-11874.

\bibitem{Roy2013_VSCFReview}
Roy,~T.~K.;\ \ Gerber,~R.~B.  {Vibrational self-consistent field calculations
  for spectroscopy of biological molecules: new algorithmic developments and
  applications},  \textit{Phys. Chem. Chem. Phys.} \textbf{2013,} \textsl{15,}
  9468-9492.

\bibitem{Arnim1999_GNIC}
von Arnim,~M.;\ \ Ahlrichs,~R.  {Geometry optimization in generalized natural
  internal coordinates},  \textit{J. Chem. Phys.} \textbf{1999,} \textsl{111,}
  9183-9190.

\bibitem{Carbonniere2010_VCIP}
Carbonni{\`e}re,~P.;\ \ Dargelos,~A.;\ \ Pouchan,~C.  {The VCI-P code: an
  iterative variation--perturbation scheme for efficient computations of
  anharmonic vibrational levels and IR intensities of polyatomic molecules},
  \textit{Theor. Chem. Acc.} \textbf{2010,} \textsl{125,} 543--554.

\bibitem{Carbonniere2012_VCIP}
Carbonni{\`e}re,~P.;\ \ Pouchan,~C.  {Modelization of vibrational spectra
  beyond the harmonic approximation from an iterative variation--perturbation
  scheme: the four conformers of the glycolaldehyde},  \textit{Theor. Chem.
  Acc.} \textbf{2012,} \textsl{131,} 1183.

\bibitem{shar14}
Sharma,~S.;\ \   {Communication: A flexible multi-reference
  perturbation theory by minimizing the Hylleraas functional with matrix
  product states},  \textit{J. Chem. Phys.} \textbf{2014,} \textsl{141,}
  111101.

\bibitem{ren16}
Ren,~J.;\ \ Yi,~Y.;\ \ Shuai,~Z.  {Inner Space Perturbation Theory in Matrix
  Product States: Replacing Expensive Iterative Diagonalization},  \textit{J.
  Chem. Theory Comput.} \textbf{2016,} \textsl{12,} 4871-4878.

\end{thebibliography}
\providecommand{\refin}[1]{\\ \textbf{Referenced in:} #1}

\end{document}